\documentclass[11pt]{report}
\usepackage{geometry}                
\usepackage{graphicx}
\usepackage{amssymb}
\usepackage{epstopdf}
\begin{document}
\begin{large}
\begin{center}
  {\Large \bf Proposal; Precise measurements of very forward particle production at RHIC}
  \vskip 10mm
  \vskip 20mm

{\large Y.Itow, H.Menjo, T.Sako, N.Sakurai} \vskip 3mm  
{\normalsize Solar-Terrestrial Environment Laboratoy / Kobayashi-Maskawa Institute for the Origin of Particles and the Universe / Graduate School of Science, Nagoya University, Japan} \vskip 10mm

{\large K.Kasahara, T.Suzuki, S.Torii}  \vskip 3mm 
{\normalsize Waseda University, Japan}  \vskip 10mm

{\large O.Adriani, L.Bonechi, R.D'Alessandro, G.Mitsuka, A.Tricomi}  \vskip 3mm
{\normalsize INFN/University of Firenze/University of Catania, Italy}  \vskip 10mm

{\large Y.Goto}  \vskip 3mm
{\normalsize RIKEN Nishina Center / RIKEN BNL Research Center, Japan}  \vskip 10mm

{\large K.Tanida}  \vskip 3mm
{\normalsize Seoul National University}  \vskip 10mm

{\large}  \vskip 3mm
{\normalsize}  \vskip 10mm

{\large}  \vskip 3mm
{\normalsize}  \vskip 10mm
\end{center}

\clearpage
\begin{center}
  {\bf ABSTRACT}
\end{center}
We propose a new experiment Relativistic Heavy Ion Collider forward (RHICf) for the precise measurements of very 
forward particle production at RHIC.
The proposal is to install the LHCf Arm2 detector in the North side of the ZDC installation slot at the PHENIX interaction
point.
By  installing high-resolution electromagnetic calorimeters at this location we can measure
the spectra of photons, neutrons and $\pi^{0}$ at pseudorapidity $\eta>$6.

The new measurements at 510\,GeV p+p collisions contribute to improve the hadronic interaction models
used in the cosmic-ray air shower simulations.
Using a kinematic coverage at RHIC similar to that of the measurements at LHC 7--14\,TeV p+p collisions,
we can test the scaling of forward particle spectra with a wide $\sqrt{s}$ range and make the extrapolation of models
into cosmic-ray energy more reliable.
Combination of a high position resolution of the RHICf detector and a high energy resolution of the ZDC 
makes it possible to determine p$_{T}$ of forward neutrons with the ever best resolution.
This enables us to study the forward neutron spin asymmetry discovered at RHIC in more detail. 
Combined data taking and analysis between PHENIX and RHICf will make it possible to identify the origin 
of each forward particle, particularly diffractive and non-diffractive interactions.

We request 510\,GeV p+p collisions with $\beta^{*}$=10\,m.
With 100 and 20 colliding and non-colliding bunches, respectively, and nominal bunch intensity 
2$\times$10$^{11}$, we can expect an instantaneous luminosity of 1.1$\times$10$^{31}$\,cm$^{-2}$s$^{-1}$.
To study the asymmetry of forward particle production, we require radially polarized beams
with a moderate polarization, 0.4-0.5 or higher.
These beam conditions provide sufficient event statistics after one day of operation.
Including another day for contingency and 1-5 days for beam setup depending on the previous beam mode,
we propose 3-7 days of beam time to complete our proposal. 

RHICf is also interested in participating a possible measurements at p+A collisions 
to understand the interaction between cosmic-ray particles and atmosphere.
Collision of light ions like nitrogen is the ultimate goal for the cosmic-ray physics, but the collision of heavy ion
is also of interest.

Our proposal is to bring the LHCf Arm2 detector to RHIC after the LHC 13\,TeV p+p collision runs 
planned  in early 2015, and then operate in RUN16 at RHIC.
Installation of the cables and a structure to fix the detector, preparation for the data acquisition especially 
synchronization with PHENIX and accelerator will be carried out in advance as soon as the proposal is approved.

\clearpage
\tableofcontents
\clearpage

\chapter{Introduction and physics} \label{sec-introduction}
\section{Cosmic-ray physics and $\sqrt{s}$ dependence of hadronic interaction}
The origin of cosmic rays is a century-standing problem.
Recent observations of ultra-high-energy cosmic rays (UHECR) by the Pierre Auger Observatory (PAO) \cite{PAO}
and Telescope Array \cite{TA} have been dramatically improved in both the statistics and the systematics.
The existence of a spectral cutoff at approximately 10$^{19.5}$\,eV is now clearly identified.
However, the interpretation of the observed results is not settled.
One of the main reasons for the difficulty is the uncertainty in air shower modeling.
Fig.\ref{fig-PAO-composition} shows the so-called X$_{max}$ parameter as a function of the cosmic-ray energy
observed by the PAO.
Here, X$_{max}$ is the height of the shower maximum measured from the top of the atmosphere in g/cm$^{2}$.
Experimental data are compared with the predictions by air shower simulation with the two extreme assumptions
that all cosmic rays are protons or iron nuclei.
Four lines in each assumption are due to the use of different interaction models in the air shower simulation.
Because of the fact that the difference between models is larger than the experimental errors,
the determination of the primary chemical composition is difficult, and hence, the nature
of the spectral cutoff is not concluded.
The determination of the chemical composition at 10$^{17}$\,eV is also important because, at approximately
this energy, the source of the cosmic rays is believed to switch from galactic to extragalactic and the chemical 
composition rapidly changes with energy \cite{KASCADE}.
However, due to the uncertainty in air shower modeling, the determination of the chemical composition at this
energy range is still model dependent. 
To solve the origin of mysterious UHECRs and to confirm the standard scenario of the
cosmic-ray origin, constraints from the accelerator experiments are indispensable.

  \begin{figure} 
  \begin{center}
  \includegraphics[width=10cm]{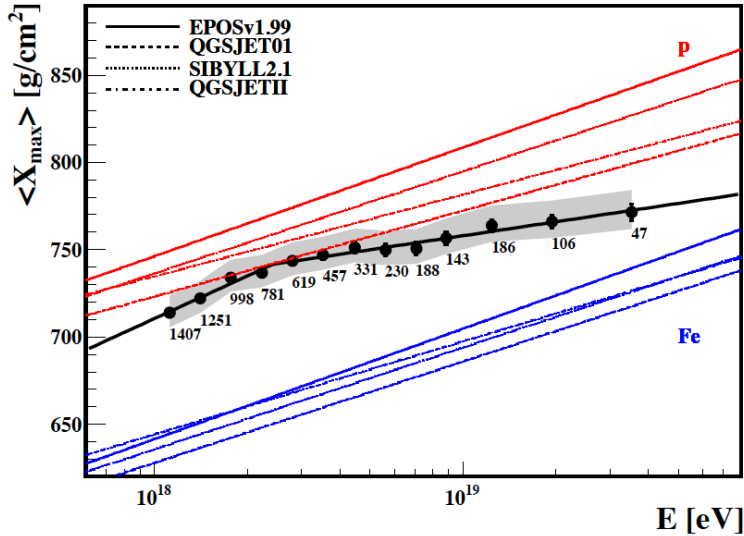}
  \vspace*{8pt}
  \caption{X$_{max}$ of air showers observed by the Pierre Auger Observatory. \cite{PAO}
  \label{fig-PAO-composition}}
  \end{center}
  \end{figure}

The difficulty in modeling hadronic interactions, which is essential to determine the air shower development,
is due to the difficulty in modeling the soft interaction described by non-perturbative QCD.
Experimentally, particles produced in such processes have a large energy flux in the forward direction and are 
difficult to measure especially in collider experiments.
Cosmic-ray interaction models have been tested with a variety of accelerator experiments with a limited
number of forward measurements, and most of the data thus far are limited in proton-proton (or anti-proton) collisions.
The Large Hadron Collider (LHC) provides an unprecedented quality of test data; this facility gives the highest
collision energy, and the experiments cover a very wide range of pseudorapidity ($\eta$) \cite{dEntteria}.
The same quality of lower-energy collision data is very important to test the $\sqrt{s}$ dependence of hadronic 
interactions to extrapolate the models beyond the LHC energy.
Fig.\ref{fig-pi0-edependence} shows the energy spectra of all $\pi^{0}$ at $\sqrt{s}$= 500\,GeV, 7\,TeV and 
50\,TeV (E$_{lab} = $1.3$\times$10$^{18}$\,eV) predicted by the DPMJET3 \cite{DPM} and QGSJET-II \cite{QGS} models.
With the particle energy scaled by the beam energy ($x_{F}$), the DPMJET3 model assumes 
a perfect scaling, whereas QGSJET-II shows a softening in higher energy collisions.
The assumption of the scaling or collision energy dependence is an important issue to be tested at 
the collider experiments.

In contrast to p+p collisions, only d+Au collisions at RHIC and  p+Pb collisions at LHC have provided collision situations
that are similar to cosmic-ray protons interacting with the atmosphere.
In both cases, strong nuclear effects were reported by STAR \cite{STAR} and ALICE \cite{ALICE}.
These effects will be important inputs to simulate proton-atmosphere collisions at extreme conditions.
However, no direct tests of nuclear effects in proton-atmosphere collisions have been performed thus far.

  \begin{figure}
  \begin{center}
  \includegraphics[width=9cm]{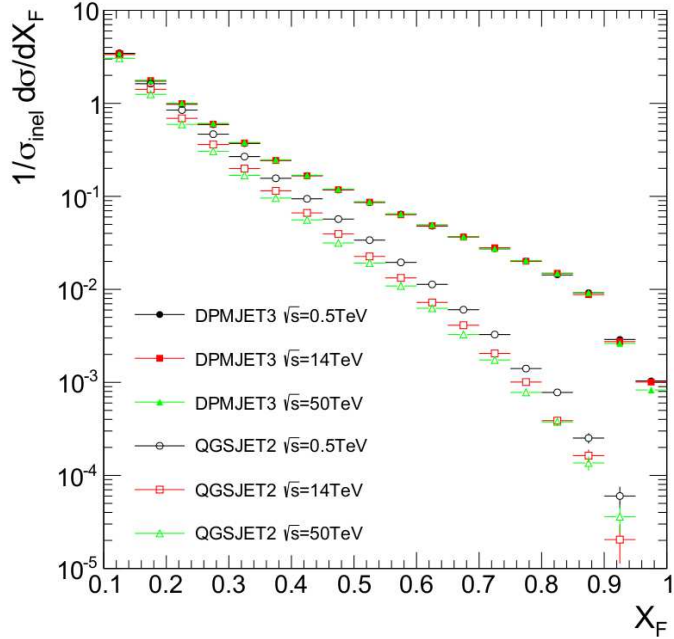}
  \vspace*{8pt}
  \caption{X$_{F}$ spectra of all $\pi^{0}$ at $\sqrt{s}$ = 500\,GeV, 7\,TeV and 50\,TeV p+p collisions.
  \label{fig-pi0-edependence}}
  \end{center}
  \end{figure}

\section{Asymmetry in the forward particle production in the polarized p+p collisions}
With the first polarized p+p collisions at $\sqrt{s}$ = 200\,GeV at RHIC, 
a large single transverse-spin asymmetry ($A_N$) for neutron production in very forward kinematics was discovered 
by a polarimeter development experiment \cite{Fukao:2006vd}.
This was recently confirmed by the PHENIX experiment using the Zero Degree Calorimeter (ZDC) and the Shower 
Maximum Detector (SMD) installed between the ZDC modules \cite{ref-PHENIX-forward}.
The discovery of the large $A_N$ for neutron production is new, 
important information to understand the production mechanism of the very 
forward neutron. 
 
  \begin{figure}[htbp]
  \begin{center}
  \includegraphics[width=7.2cm]{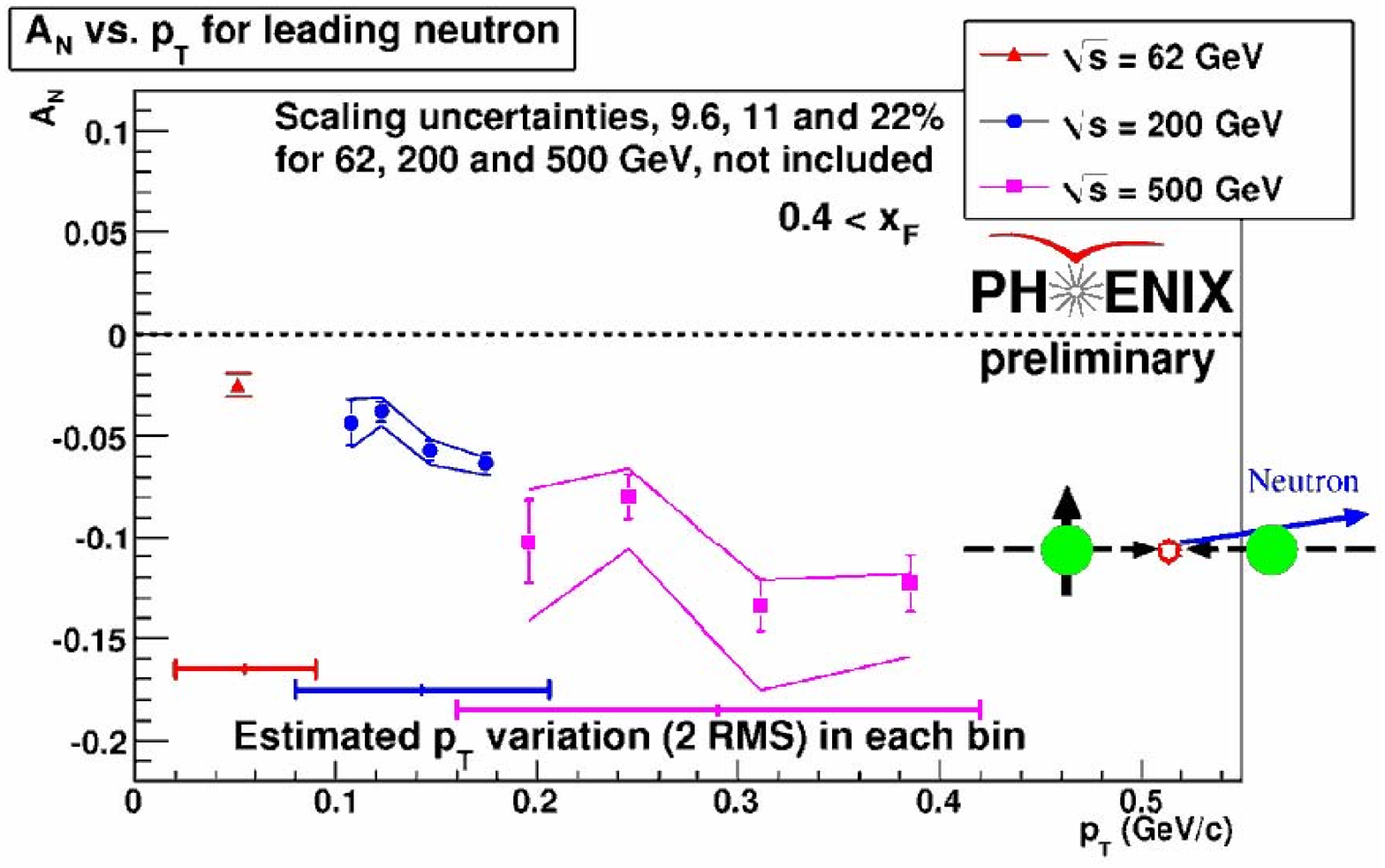}
  \includegraphics[width=7.2cm]{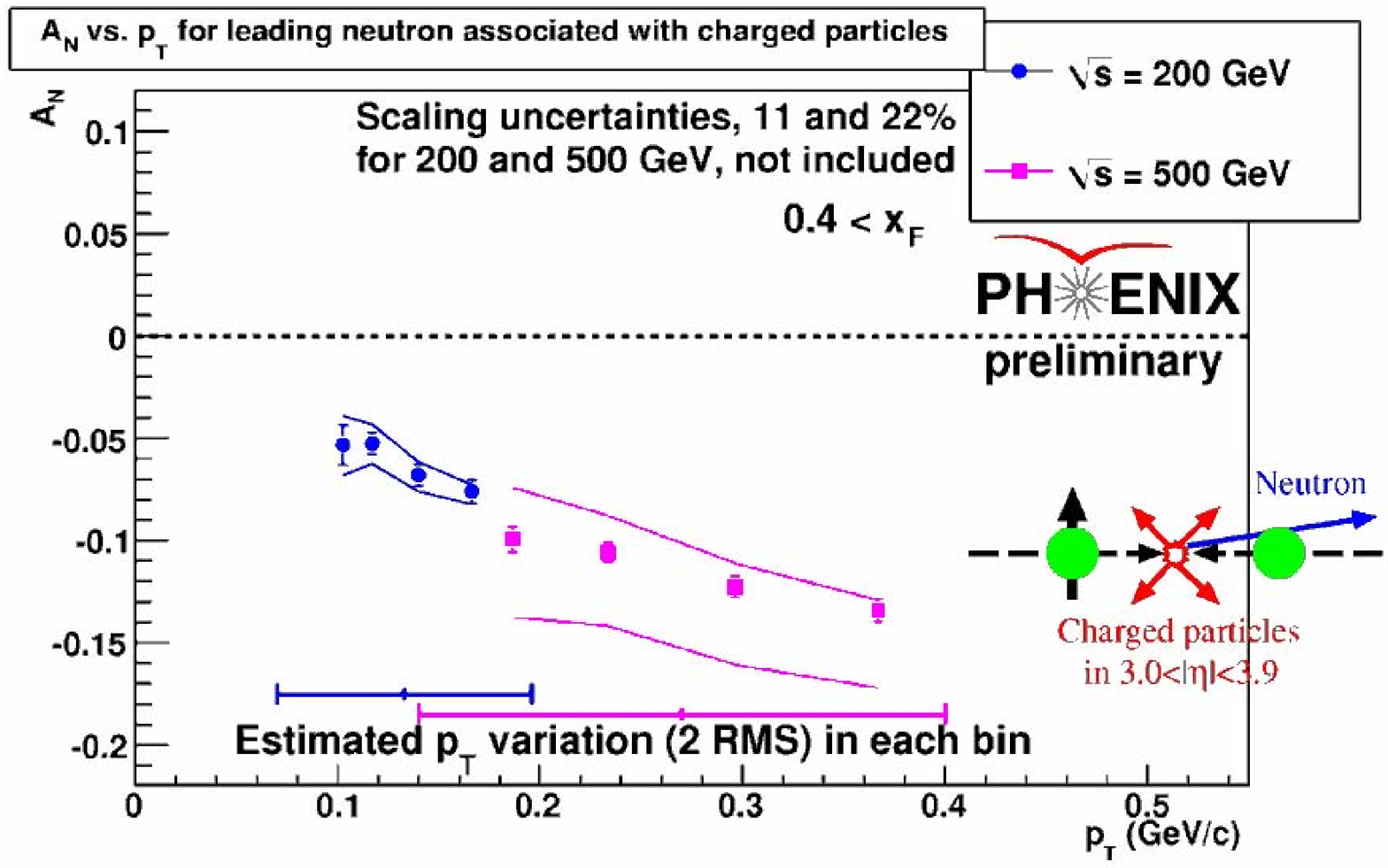}
  \caption{The measured asymmetries of very forward neutron production as 
  functions of $p_T$ with an inclusive-neutron trigger (left) and with 
  a semi-inclusive neutron trigger including a beam-beam interaction 
  requirement (right).}
  \label{fig:asym_pt}
  \end{center} 
  \end{figure}

The $\sqrt{s}$ dependence of $A_N$ from three different collision energies, 62.4\,GeV, 200\,GeV, and 
500\,GeV was studied \cite{ref-PHENIX-forward-tanida}. 
The result is shown in Fig.~\ref{fig:asym_pt}. 
The hit position dependence on the detector was measured at each energy, 
although this dependence was largely smeared by the position resolution. 
The result was converted to the $p_T$ dependence, which showed a hint of the $p_T$ scaling property 
of $A_N$ of the very forward neutron production. 
The asymmetry is caused by interference between spin-flip and non-flip 
amplitudes with a relative phase. 
Kopeliovich et al. \cite{Kopeliovich:2011bx} studied the interference of a pion and $a_1$, or a pion and $\rho$ 
in the $1^+S$ state. 
The data agreed well with the independence of energy. 
The asymmetry is sensitive to the presence of different mechanisms, e.g., Reggeon exchange with 
spin-non-flip amplitudes, in addition to the dominant one-pion exchange amplitude, even if these amplitudes are small. 

\section{LHCf experiment}
Large Hadron Collider forward (LHCf) is one of the LHC experiments to measure forward neutral particles to 
calibrate the interaction models used in the cosmic-ray physics \cite{LHCf}.
LHCf successfully acquired to take data for LHC 900\,GeV, 2.76\,TeV and 7\,TeV
p+p collisions and 5.0\,TeV ($\sqrt{s_{NN}}$) p+Pb collisions.
LHCf installed compact calorimeters at the installation slots of the ZDCs located 140\,m
from an interaction point of the LHC.
Two independent detectors called Arm1 and Arm2 at either side of the interaction point were installed.
At this location, neutral particles (predominantly photons decayed from $\pi^{0}$ and neutrons) emitted at
$\eta>$8.4 are observed.
Each detector has two small calorimeter towers that allow simultaneous detection of two high-energy particles
and hence identification of photon pairs originating from $\pi^{0}$ by reconstructing the invariant mass of these
particles.
The performance of the detectors at LHC was satisfactory and well understood \cite{LHCf-performance}.
LHCf is preparing for measurements at LHC 13\,TeV p+p collisions in April to May, 2015.

Thus far, LHCf has published energy spectra of forward photons at 900\,GeV \cite{LHCf900GeV} and 7\,TeV \cite{LHCf7TeV} 
and forward $\pi^{0}$ spectra at 7\,TeV \cite{LHCfpi0}.
The suppression of forward $\pi^{0}$ production in the p+Pb collisions was also reported recently \cite{LHCf-pPb}.
As mentioned above, comparisons of data at different collision energies are extremely important.
Fig.\ref{fig-photon-coverage} shows a MC prediction of forward photon yield ($\frac{d^{2}N}  {N_{0}~dx_{F}dp_{T}}$).
In a model characterized by perfect scaling this distribution is $\sqrt{s}$ independent.
The triangles in the figure indicate the phase-space coverages with LHC set at 900\,GeV and 7\,TeV, and also
with RHIC set at 200\,GeV and 500\,GeV.
It is found that the LHC 7\,TeV and RHIC 500\,GeV settings cover an almost identical phase space, while
the other operating points have different and very limited phase-space coverages.
The prime motivation for RHICf is to measure the forward particles production with identical phase-space coverages and
test the $\sqrt{s}$ scaling.

  \begin{figure}[h]
  \begin{center}
  \includegraphics[width=8cm]{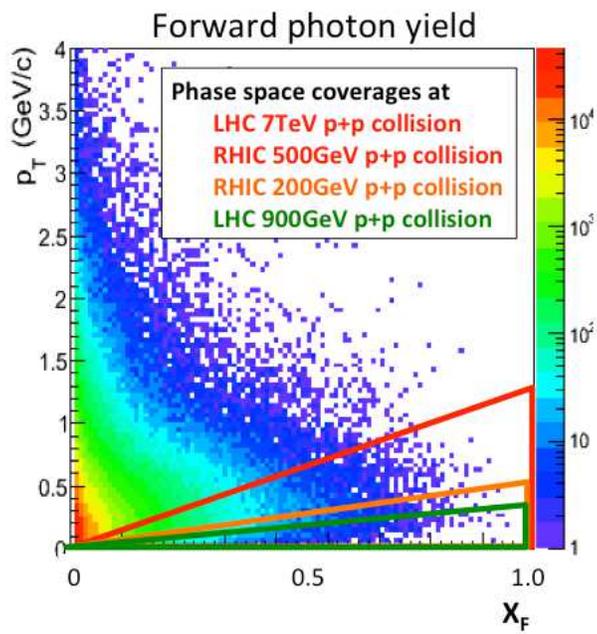}
  \caption{Comparison of phase space coverages of photons between LHCf and RHICf.}
  \label{fig-photon-coverage}
  \end{center} 
  \end{figure}

\chapter{Experimental setup and detector} \label{sec-setup}
\section{LHCf detector at RHIC}

The idea at the basis of the RHICf proposal is to bring one of the LHCf detectors \cite{LHCf-JINST} \cite{LHCf-silicon}
called Arm2 shown in Fig.\ref{fig-RHICf-schematic}
to RHIC and to install it in front of a ZDC at the interaction point of the PHENIX experiment.
The LHCf Arm2 detector, called the RHICf detector hereafter, is composed of two sampling calorimeter towers
with transverse dimensions to the beam of 25\,mm$\times$25\,mm and 32\,mm$\times$32\,mm.
Each tower is composed of 44 radiation lengths of Tungsten interleaved with 16 scintillator layers.
Signals from scintillators are measured by 32 PMTs (HAMAMATSU R7400U). 
Eight layers (4 XY pairs) of silicon strip sensors are inserted to measure the lateral distribution of the showers.
The calorimeters have an energy and a position resolutions of $8\%/\sqrt{E/100\,GeV}+1\%$ and $<$150\,$\mu$m ($>$50\,GeV), 
respectively, for electromagnetic showers in the LHC environment.
The detector is contained in an aluminum box with the dimensions of 92\,mm (transverse to the beam direction) $\times$ 
290\,mm (along the beam) $\times$ 620\,mm in height.  
This can fit the 100\,mm gap between the beam pipes located at 18\,m from the interaction point of RHIC where ZDCs
are located. 
By installing the RHICf detector in front of a ZDC at the PHENIX interaction point, RHICf measures the neutral 
particles around zero degree.
The location of the installation is supposed at the North side of PHENIX due to the available space at the time of May 2014
and to keep a possibility of observing from the proton remnant side in p+A collisions.\\

  \begin{figure} [h]
  \begin{center}
  \includegraphics[width=10cm]{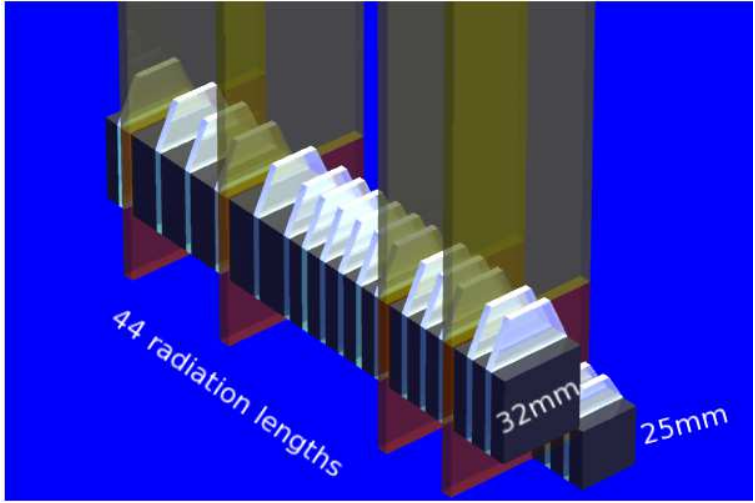} 
  \vspace*{8pt}
  \caption{Schematic view of the RHICf detector.
  \label{fig-RHICf-schematic}}
  \end{center}
  \end{figure}

  \begin{figure}[h]
  \begin{center}
  \includegraphics[width=7.5cm]{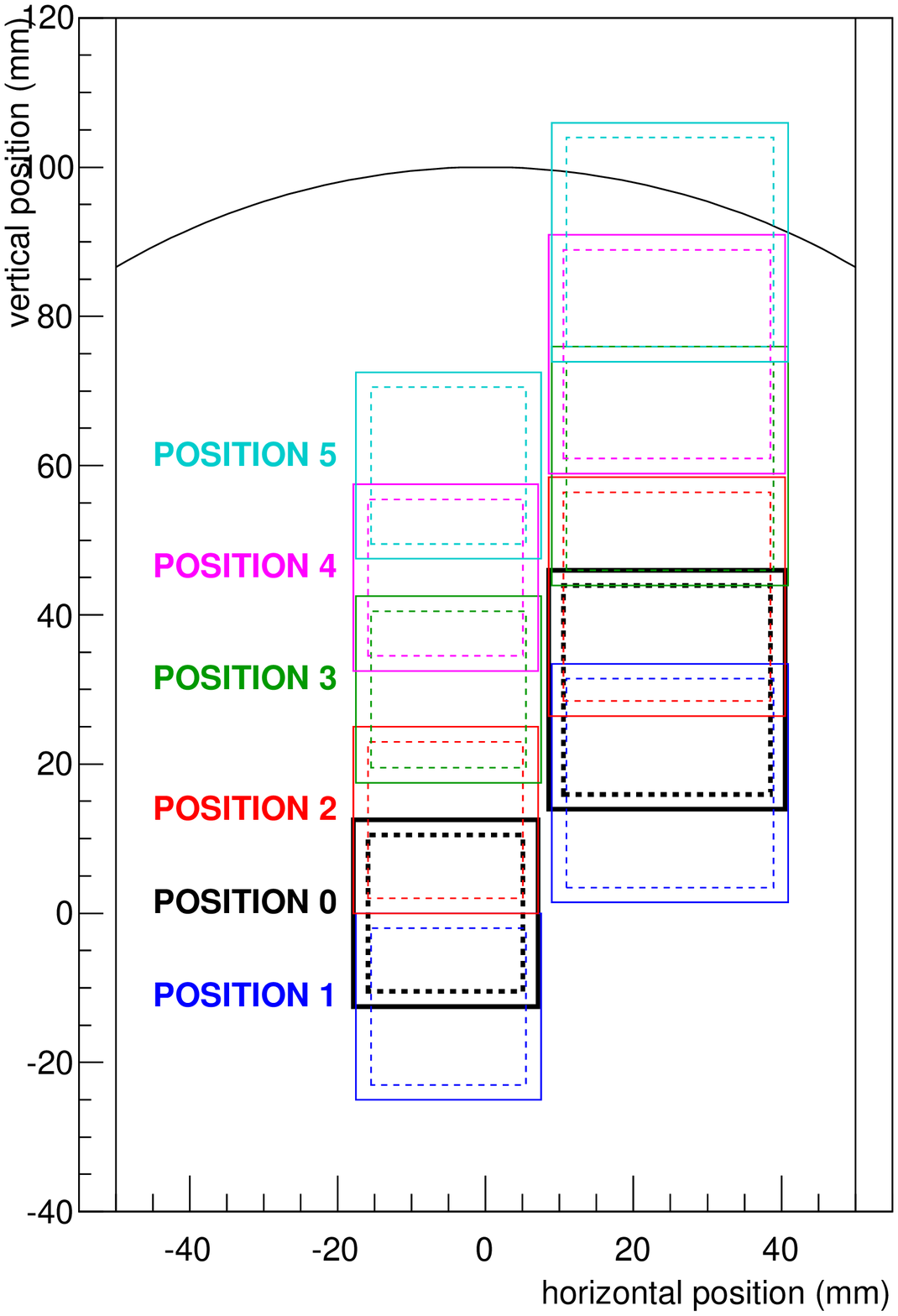}
  \caption{Candidate scan positions.  The origin is defined at the zero degree or $\eta$=$\infty$. The arc at the top indicates the observation limit, $\eta$=6, constrained by the narrowest beam pipe.  Small and large squares indicate the coverage of the 25\,mm and 32\,mm calorimeters, respectively.  Dashed squares indicate the fiducial area used in the standard LHCf analysis.
  \label{fig-scanpos}}
  \end{center} 
  \end{figure}

The RHICf detector will be suspended from a structure called manipulator that allows a vertical movement of the detector.
The purposes of the manipulator are 1) to increase the rapidity (or p$_{T}$) coverage, 2) to avoid
interference to ZDC running when RHICf is not in operation.
At the location of ZDC neutral particles are detectable up to the radius r=100\,mm or $\eta$=6.
On the other hand, because of the transverse sizes of the RHICf calorimeters, the accessible phase space is not
covered all at once.
A possible pattern of the vertical scan positions, their coverage in the p$_{T}$-x$_{F}$ plane and acceptance for
different categories of events are summarized in Fig.\ref{fig-scanpos}, Fig.\ref{fig-coverage} and Tab.\ref{tab-acceptance}. 
Acceptance is defined as an expected number of events in an inelastic collision calculated based on the generator 
PYTHIA 8.185 \cite{PYTHIA}.

  \begin{figure}[h]
  \begin{center}
  \includegraphics[width=11.8cm]{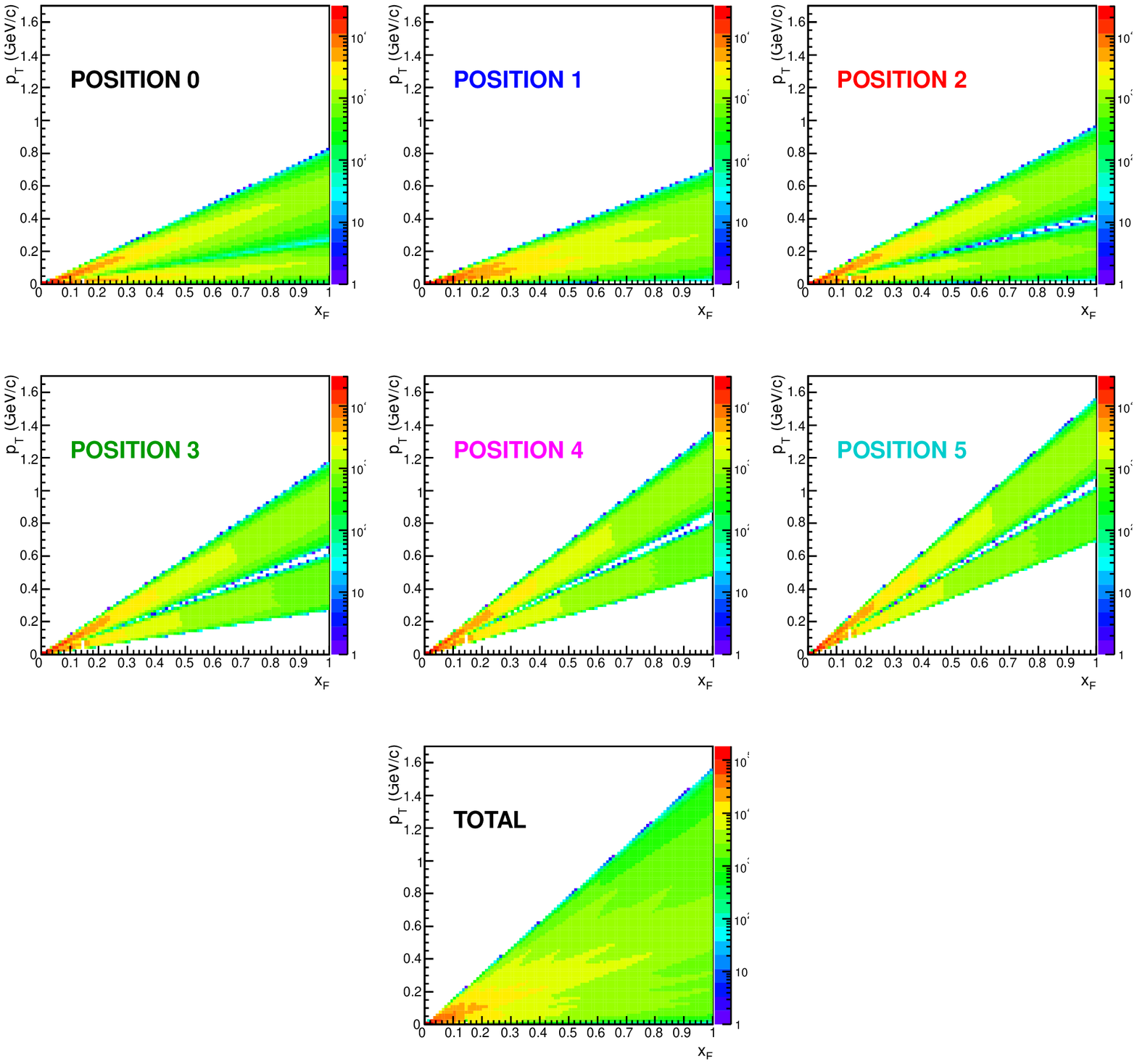}
  \caption{Coverage in the p$_{T}$-x$_{F}$ plane for 6 scanning positions given in the Fig.\ref{fig-scanpos} and 
  a sum of them at the bottom.   The color represents the effective fiducial area of the calorimeters
  in log scale.} 
  \label{fig-coverage}
  \end{center} 
  \end{figure}

  \begin{table}[th]
  \begin{center}
  \caption{Acceptance of RHICf detector for 6 positions and deferent event categories.  Acceptance is defined as 'event/inelastic collision' using PYTHIA 8.185.  Fiducial cut in the standard LHCf analysis is applied.  Energy thresholds of 30\,GeV and 100\% detection efficiency were assumed.  TS and TL designate Small Tower (calorimeter) and Large Tower, respectively.}
  \vskip 3mm
  \begin{tabular}{ccccccc}
                &                & \multicolumn{2}{c}{photon} & \multicolumn{2}{c}{neutron}  &$\pi^0$ \\
  \hline
  Position & All event & TS                 & TL              & TS                   & TL               &             \\
  \hline
  \hline
  0            & 0.047     &  0.0079          & 0.011         &   0.014            & 0.014          &   0.00023 \\
  1            & 0.049     &  0.0077          & 0.012         &   0.013            & 0.016          &   0.00025 \\
  2            & 0.043     &  0.0077          & 0.010         &   0.013            & 0.012          &   0.00020 \\
  3            & 0.034     &  0.0069          & 0.0087       &   0.0094          & 0.0086        &   0.00015 \\
  4            & 0.027     &  0.0060          & 0.0074       &   0.0071          & 0.0067        &   0.00011 \\
  5            & 0.019     &  0.0052          & 0.0047       &   0.0054          & 0.0039        &   0.00006 \\
  \end{tabular}
  \end{center}
  \label{tab-acceptance} 
  \end{table}

  \begin{figure}[h]
  \begin{center}
  \includegraphics[width=10.0cm]{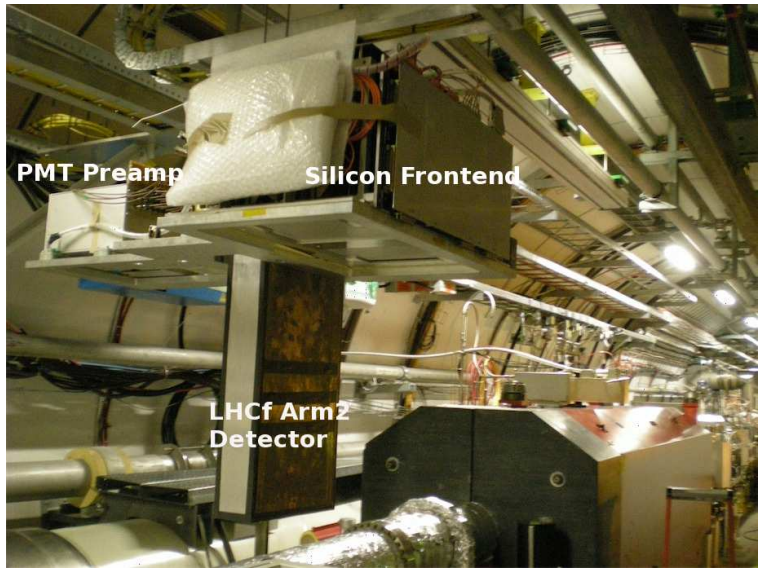}
  \caption{The LHCf Arm2 detector and its fronted electronics during the installation in the LHC tunnel.}
  \label{fig-arm2-assebly}
  \end{center} 
  \end{figure}

  \begin{figure}[h]
  \begin{center}
  \includegraphics[width=10.0cm]{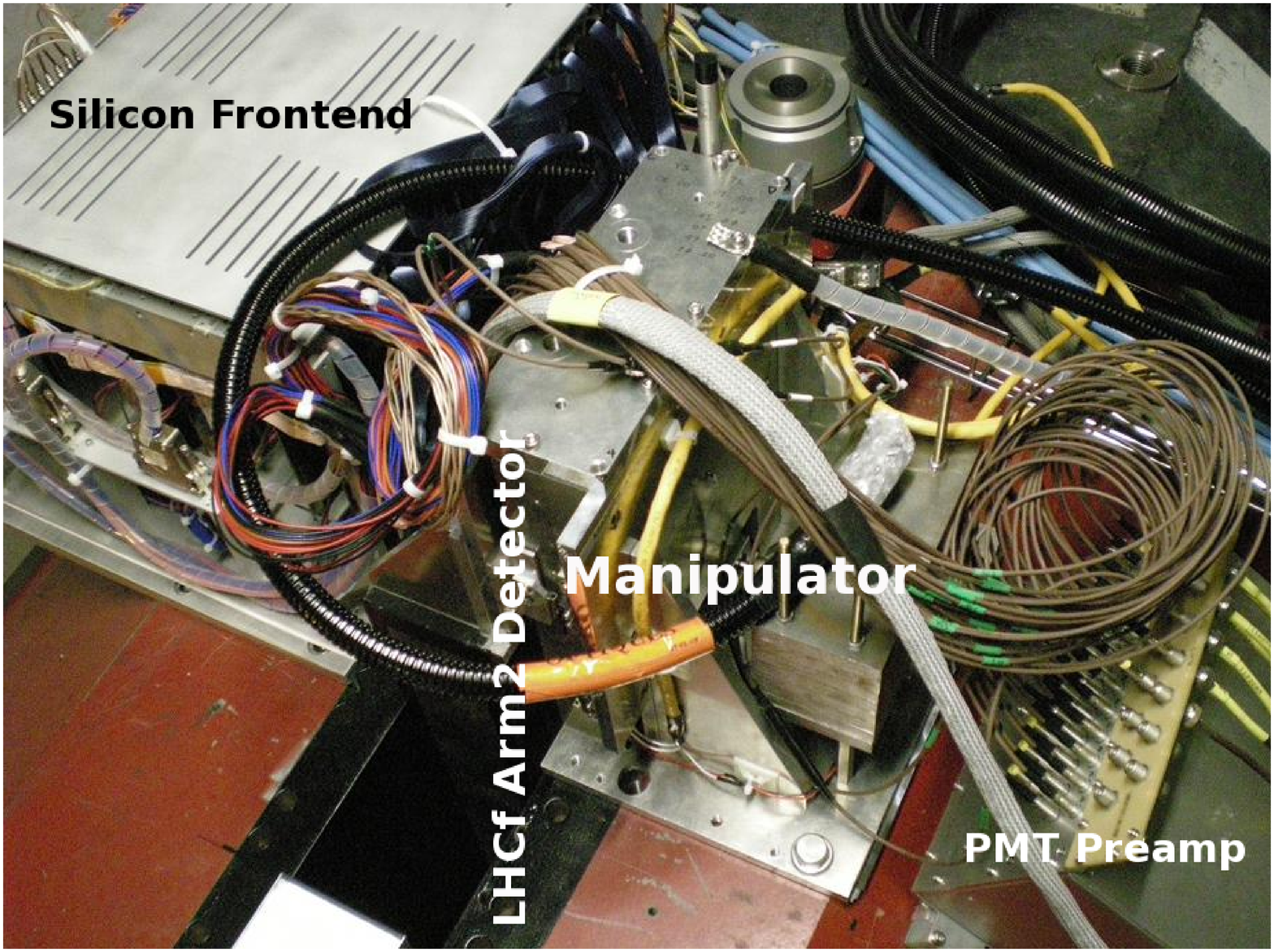}
  \caption{The LHCf Arm2 detector and its fronted electronics fixed on the TAN.}
  \label{fig-arm2-atTAN}
  \end{center} 
  \end{figure}

  \begin{figure}[h]
  \begin{center}
  \includegraphics[width=6.5cm]{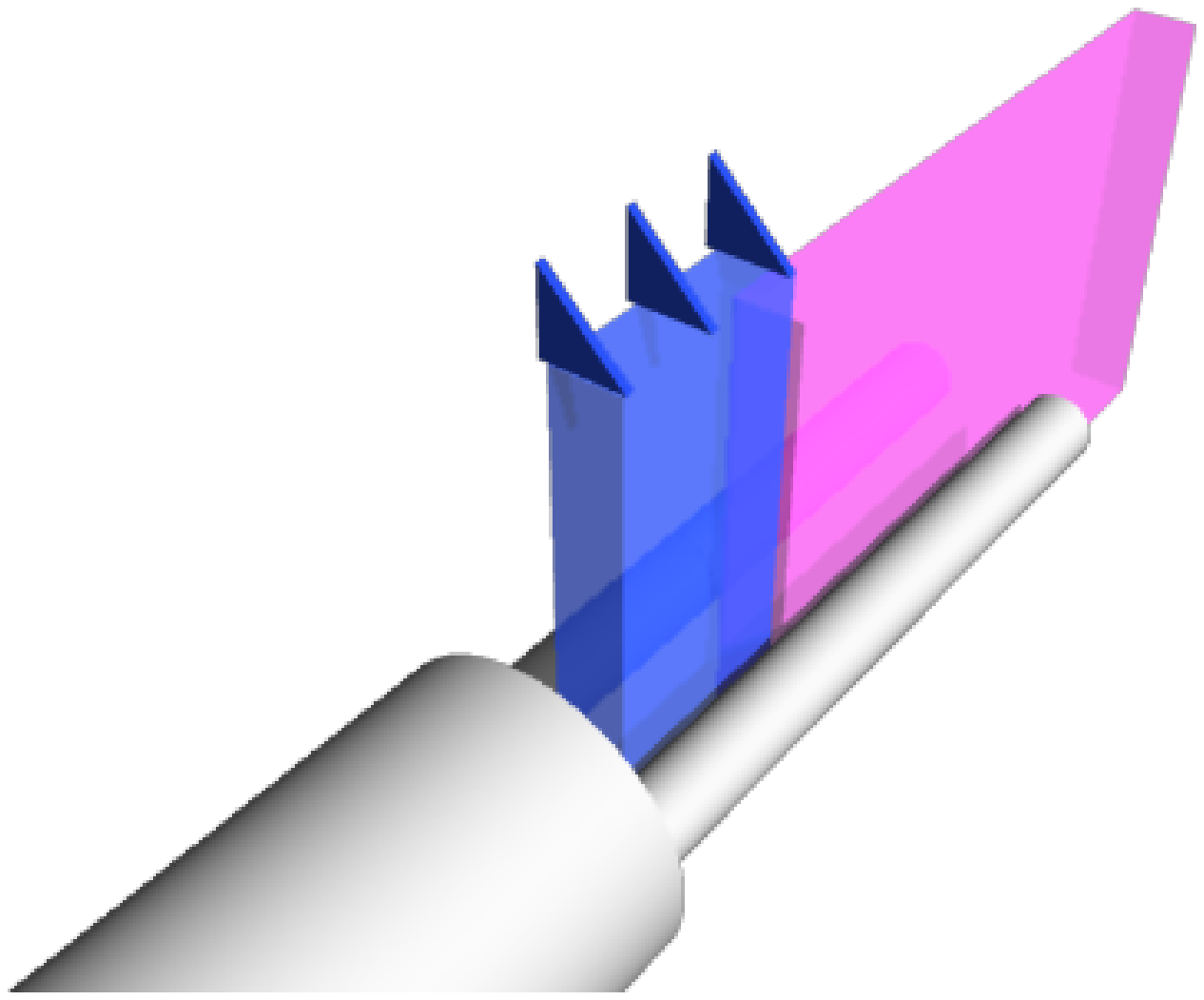}
  \includegraphics[width=6.5cm]{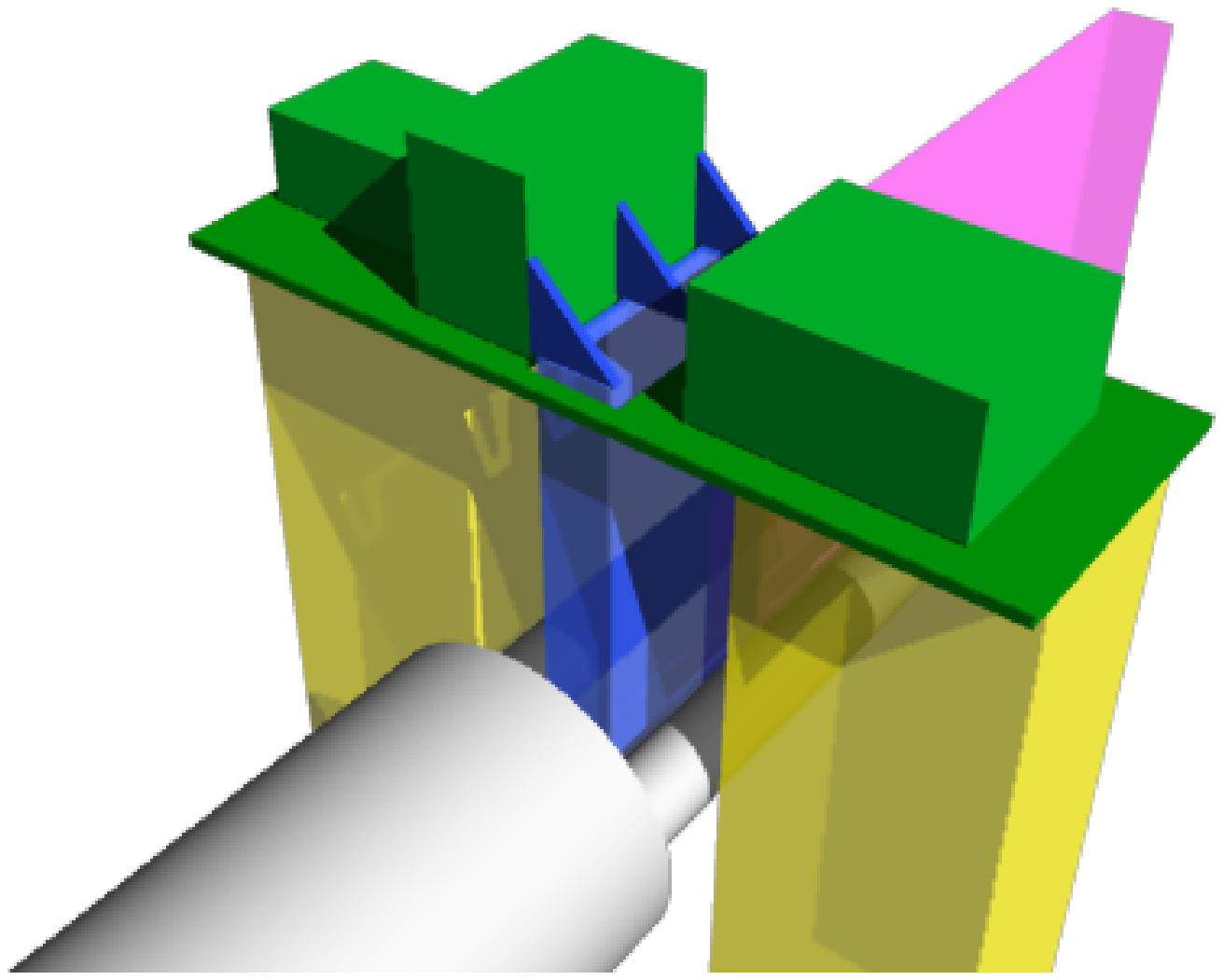}
  \caption{The LHCf Arm2 detector and its fronted electronics fixed on the TAN.}
  \label{fig-RHICf-image}
  \end{center} 
  \end{figure}

Together with the manipulator, the detector is installed with the front-end circuit of the silicon strip
sensors and the PMT preamplifiers.
All components were assembled on a single aluminum plate and installed in the LHC tunnel when
they were used in the LHCf experiment as shown in Fig.\ref{fig-arm2-assebly}.
Fig.\ref{fig-arm2-atTAN} shows a photograph when the detector and electronics were fixed on the
TAN radiation shield (red part in photo) in the LHC tunnel.
RHICf needs the same identical components except for the manipulator that must be upgraded to allow a wider movable
range at RHIC.
Conceptual images showing how the RHICf detector and its electronics will be installed in the RHIC
tunnel are shown in Fig.\ref{fig-RHICf-image}.
Left figure shows the beam pipes (grey), RHICf (blue) and ZDC (magenta) while right figure includes the
manipulator and the electronics (green) and their support structure (yellow).
It is necessary to construct the structure to hold the RHICf equipment which weighs about 80\,kg in total.

\section{Data acquisition}
RHICf records the signals from 32 PMTs for the calorimeters and from 3,072 silicon strips (3 clock samples for each strip).
The signals from the silicon strip sensors are digitized near the detector and transferred to the rack room
through optical fibers.
The 3\,$\mu$sec deep analog pipeline of the silicon readout was designed to work with the 40\,MHz LHC clock, 
but it is confirmed that it is also operational at 37.7\,MHz corresponding to 4$\times$(RHIC clock).  
On the other hand, the analog signals from PMTs are amplified and transferred to the rack room through 
coaxial cables.
The charge from the PMTs are recorded by charge ADCs (CAEN V965) through a conventional technique; linear fanout,
discriminators (CAEN V814) and analog delay cables. 
The discriminator outputs are sent to a FPGA module and a trigger signal is issued when an existence of a 
high-energy shower is identified.
A time chart of the signal processing is outlined in Fig.\ref{fig-time-chart}.
To synchronize with the collisions a signal synchronized with each bunch (spin flag signals, for example) and a clock 
signal provided by the RHIC machine are required.

  \begin{figure}[h]
  \begin{center}
  \includegraphics[width=14cm]{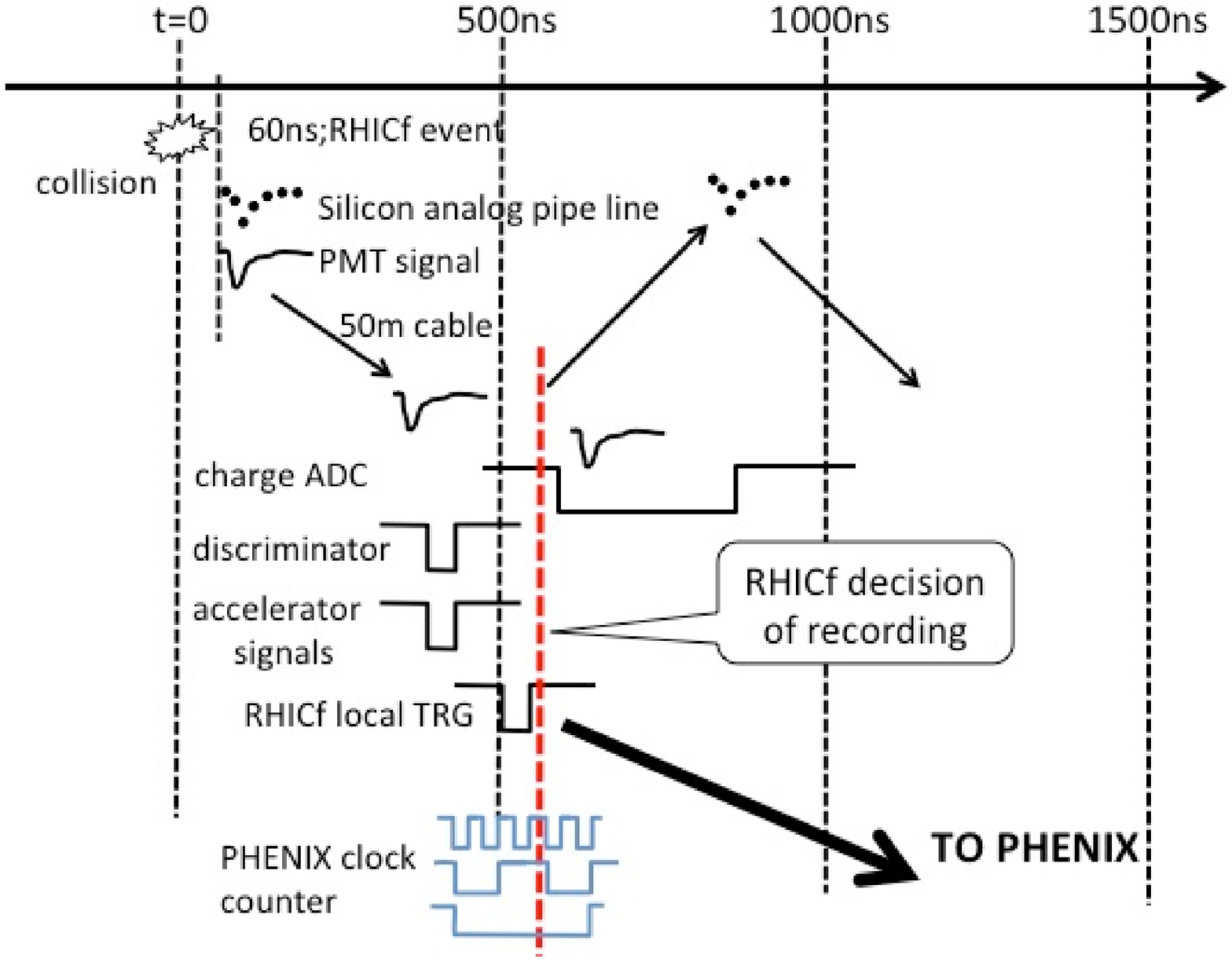}
  \caption{Time chart of the RHICf data acquisition.}
  \label{fig-time-chart}
  \end{center} 
  \end{figure}

The space for installing the electronics could be provided by the PHENIX group in the PHENIX rack room and 
also about 50 cables (30--80\,meters long) must be installed.
The length of the cables depends on whether they will follow the standard path or whether a short cut to the rack room
can be used.
Some power supplies will be installed near the detector where it is shielded from radiation and some electronics are
already in operation.
The length of the cables from the detector to the power supplies is supposed to be 10\,m.
A list of necessary cables and a schematic diagram are shown in Tab.\ref{tab-cables} and Fig.\ref{fig-cabling}. 
More detail investigation of the cabling will be discussed in 2014 summer after RUN14.

  \begin{table}[t]
  \begin{center}
  \caption{List of cables between the detector and the rack room (long cables), between the detector and the power supplies (short cables).}
  \vskip 3mm
  \begin{tabular}{lcl}
  Type (connector)                      & Quantity     & Purpose           \\
  \hline
  \hline
  [Long cables] &&\\
  50$\Omega$ coaxial (BNC)     &  32+10   & PMT signal, encoder         \\
  25 wires low current (DSUB)    &  3            & Sensors (ex. T mon) \\
  Optical fiber bundle                   &  2     & Silicon control\\
  (16 single mode; custom made)  &                &\\
  Optical fiber (multi mode)         &  6            & Silicon readout \\
  Optical fiber                              &  1            & PMT calibration \\
  \hline
  [Short cables] &&\\
  37 wires HV (1000V; REDEL)  &  1               & PMT HV           \\
  16 wires MV (100V)                 &  1               & Silicon bias      \\
   2 wires high current ($<$10A)     &16               & Silicon frontend\\
   3 wires mid current ($<$3A)        &   1              & Manipulator drive\\
   2 wires low current ($<$1A)        & 16               & Silicon FE sensors \\
   16 wires low current ($<$1A)      & 1                & Silicon bias sensors\\
  
  \end{tabular}
  \end{center}
  \label{tab-cables} 
  \end{table}

  \begin{figure}[h]
  \begin{center}
  \includegraphics[width=14cm]{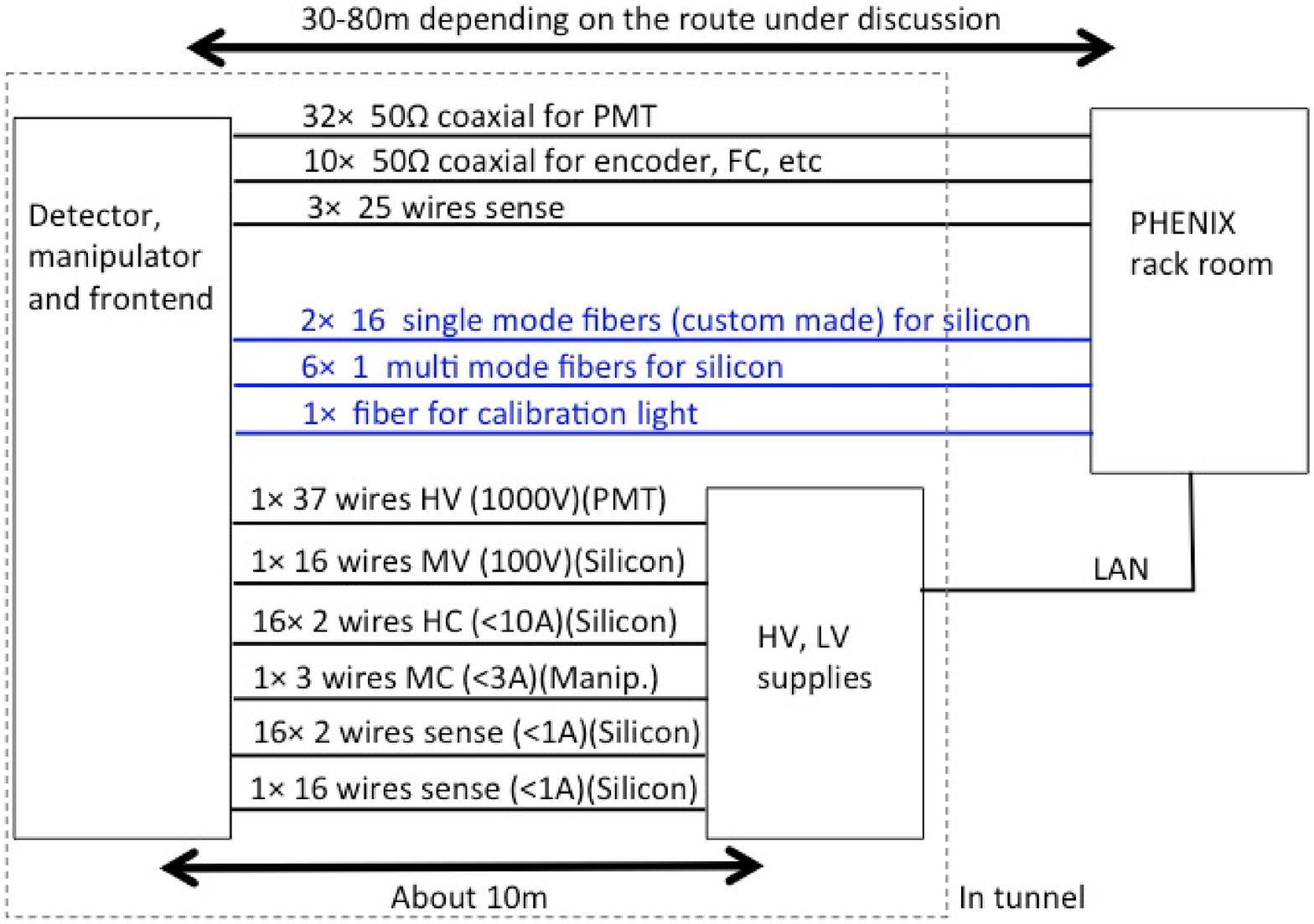}
  \caption{Plan of cabling between the RHICf detector and the PHENIX rack room.}
  \label{fig-cabling}
  \end{center} 
  \end{figure}

\hskip 5mm

\noindent {\bf Constraints in the event rate} \\
The data recording in the rack room is carried out through VME bus.
This limits the speed of the RHICf data acquisition to 1\,kHz and results in a large inefficiency at high luminosity.
Not only the inefficiency but signal overlap causes problems at high luminosity.
The RHICf calorimeters use Gd$_{2}$SiO$_{5}$ (GSO) scintillators to avoid radiation damage during LHC operation.
The slow decay constant of the GSO scintillators was also
important to avoid a saturation of the PMTs at LHC.
The slow decay constant, however, causes an overlap of signals when successive events occur within a
short time interval.
For safety, 10\,$\mu$sec is assumed when we discuss about the beam condition and operation plan in 
Chapter.\ref{sec-require}.    

\section{Joint operation with PHENIX}
Recording the events from identical collisions between PHENIX and RHICf will open a new possibility for the
collider experiments.
1) This joint analysis enables a similar analysis to the one carried out by PHENIX ZDC+SMD and BBC for
spin asymmetry measurement.
2) As discussed in our LOI \cite{RHICf-LOI}, joint analysis of RHICf and PHENIX ZDC will improve the
p$_{T}$ resolution for forward neutrons.
3) Correlation between the very forward detector and the central detector will provide new information
about low mass diffraction events.
4) Should RHICf can operate during p+A collisions, information from the central detector is useful to estimate
impact parameter of each collision.
5) In p+A collisions, information from the central detector is useful to extract the events from Ultra Peripheral 
Collisions that was observed by LHCf in the p+Pb collisions \cite{LHCf-pPb}.

The current idea of PHENIX-RHICf joint operation is to trigger PHENIX according to the RHICf trigger signals.
As shown in the time chart in Fig.\ref{fig-time-chart}, RHICf can decide to record data after about 500\,ns
from a collision.
Using this signal, PHENIX has enough time to record the data for the same event.
Because the maximum PHENIX DAQ speed is 5\,kHz while it is 1\,kHz for RHICf, PHENIX is
in principle able to record all of the RHICf triggered events.
Usage of PHENIX trigger signals for the RHICf trigger is also under consideration.
To ensure offline event identification between two experiments, RHICf will record the clock counter signal
provided by PHENIX.
This counter is an accumulated count of 10\,MHz clock with a 64-bit dynamic range, thus provides an
absolute time for each event.  

\chapter{Beam conditions and operation scenario} \label{sec-require}
\section{Beam conditions and setup time}
To obtain a wide p$_{T}$ coverage comparable to the LHCf measurements, operation at 510\,GeV p+p collisions
is our target.  
To avoid an angular divergence, a high $\beta^{*}$ is needed.
Assuming a nominal emittance $\epsilon^{*}$=20\,mm\,mrad and $\beta^{*}$=10\,m, 
86\,$\mu$rad of angular dispersion is expected.
This results in a 1.5\,mm dispersion at the RHICf distance which is compatible with our physics goal. 

Due to the high $\beta^{*}$, luminosity becomes reasonably low even with a nominal bunch intensity 
I$_{b}$=2$\times$10$^{11}$.
With 100 collision bunches the total luminosity and collision
pileup ($\mu$) are 1.1$\times$10$^{31}$\,cm$^{-2}$s$^{-1}$ and 6.5\%, respectively.
Insertion of about 20 non-colliding bunches is necessary to study the background from beam halo and interaction
between the beam and residual gas in the vacuum chamber.
Considering the maximum RHICf acceptance, 0.05, in Tab.\ref{tab-acceptance}, and the inelastic cross
section $\sigma_{ine}$=50\,mb, the RHICf event rate will be 28\,kHz though the recording rate will be limited to 1\,kHz 
because of the aforementioned DAQ limits.
Prescaling of the single calorimeter events down to $<$1\,kHz while recording all double calorimeter events will be
applied to enhance the relatively low rate of $\pi^{0}$ events (140\,Hz at position 1).
Also enhanced triggers for high energy showers and a rough classification between electromagnetic and hadronic
showers are possible.
According to the largest acceptance of a single calorimeter (0.012+0.016 for the Large Tower at the position 1 in 
Tab.\ref{tab-acceptance},), the fraction of multiple shower events in a single calorimeter due to the collision pileup 
is 0.2\% (0.065$\times$0.028) at maximum and is at an acceptable level.
The probability that successive events occur within 10\,$\mu$sec is 15\%.
These events may deteriorate the charge measurement of the calorimeters due to the slow GSO decay
time, but they can be identified and eliminated in the analysis.

To study single spin asymmetry of the forward particle production, we require a polarized p+p collision.
Since RHICf can perform scans by moving its position only in the vertical direction, radial polarization resulting an 
up-down asymmetry is required.
Mixture of alternating polarization directions is important to reduce systematic uncertainty.
Reasonably high polarization, 0.4--0.5, is required.
Any higher polarization is always appreciated.

The required beam parameters for the RHICf operation are summarized in Tab.\ref{tab-beamparam}.
They were also studied by the CA-D section and summarized in the `RHIC Collider Projection
(FY2014--FY2018) version 6 April 2014 \cite{RHIC-Projection}.'
According to \cite{RHIC-Projection}, the expected setup time for the RHICf requested beam conditions are
given depending on the previous beam mode. 
\begin{itemize}
  \item Previous mode: polarized protons at the same energy
    \begin{itemize}
      \item 1 day of setup is needed
      \item expected polarization is the same as in the previous running mode
    \end{itemize}
  \item Previous mode: polarized protons at different energy
    \begin{itemize}
      \item 2 days of setup are needed
      \item some reduction in the proton intensity per bunch
      \item expected polarization at 255 GeV is up to 55\%
    \end{itemize}
  \item Previous mode: heavy ions
    \begin{itemize}
      \item 4-5 days of setup are needed
      \item reduction in the proton intensity per bunch by up to 30\%
      \item expected polarization at 255 GeV is up to 50\%
      \item since the polarimeters also need commissioning time, the polarization measurements will have a large error
    \end{itemize}
\end{itemize}

\noindent
Therefore RHICf operation at the end of 510\,GeV polarized p+p collision program is strongly preferred.

  \begin{table}[t]
  \begin{center}
  \caption{Required beam parameters for 510\,GeV p+p collision.}
  \vskip 3mm
  \begin{tabular}{lc}
  Parameter                      & Value     \\
  \hline
  \hline
  Beam energy (GeV)                             &  255  \\
  Beam intensity                                     &  2$\times$10$^{11}$ \\
    (protons per bunch)                           &                                   \\
  Number of colliding bunch                   &  100                           \\
  Number of non-colliding bunch            &  20                   \\
  Beam emittance (mm~mrad)               &  20 \\
  $\beta^{*}$  (m)                                    &  10                             \\
  Luminosity (cm$^{-2}$s$^{-1}$)           &  1.1$\times$10$^{31}$\\
  Polarization direction                            & radial   \\
  Polarization amplitude                          & 0.4--0.5                      \\
  Operation time                                      & 1 day  \\
  \end{tabular}
  \end{center}
  \label{tab-beamparam} 
  \end{table}

\section{Operation scenario with 510\,GeV p+p collisions}
A schedule before RUN16 will be described in Chap.\ref{sec-schedule}.
Once beam operation starts, RHICf will wait for its operation time at the garage position defined as the highest 
position within reach of the manipulator.
Even at this position, particles produced in the collisions will anyway arrive at the detector after interacting with the beam pipe.
Using these background particles, tuning of the electronics, mainly timing synchronization with collision, will be
carried out.
No special beam condition is required in this tuning phase.

During the dedicated operation time, RHICf will change its vertical position according to Fig.\ref{fig-scanpos}.
Because this causes an unstable response of ZDC behind the RHICf detector, luminosity determination of
PHENIX must be carried out using other channels. 
As shown in Tab.\ref{tab-acceptance}, the RHICf acceptance depends on the vertical detector position.
Assuming a luminosity live-time about 8 hours, though this may be longer with $\beta^{*}$=10\,m, a few sets of 30\,min
runs for each position will be performed in a single fill (delivered luminosity is 20\,nb$^{-1}$/30\,min with a
peak luminosity).
Since DAQ speed is limited anyway at 1\,kHz, 1.4\,M events at each position
are obtained after taking account of 20\% inefficiency in the analysis.
Assuming a single shower prescaling at 800\,Hz (800\,Hz/28\,kHz=3\%) and 70\% neutron detection efficiency,
statistics in each event category are shown in Tab.\ref{tab-eventrate} according to the acceptance in Tab.\ref{tab-acceptance}.
Corresponding effective integrated luminosities (and number of inelastic collisions) 
are 0.5\,nb$^{-1}$ (2.4$\times$10$^{7}$) and 16\,nb$^{-1}$ (8$\times$10$^{8}$) for single shower events and 
$\pi^{0}$ events, respectively.
Detailed studies such as expected observed spectra and sensitivity to the asymmetry measurements with reasonable binnings are given in Chap.\ref{sec-result}.

 
   \begin{table}[h]
  \begin{center}
  \caption{Expected event statistics in 30\,min of operation at L=1.1$\times$10$^{31}$.  A prescaling of single shower events (photons and neutrons) to 800\,Hz, 70\% neutron detection efficiency and 20\% offline event reduction are assumed. TS and TL designate Small Tower (calorimeter) and Large Tower, respectively.}
  \vskip 3mm
  \begin{tabular}{ccccccc}
                &                & \multicolumn{2}{c}{photon} & \multicolumn{2}{c}{neutron}  &$\pi^0$ \\
  \hline
  Position & All event & TS                 & TL              & TS                   & TL               &             \\
  \hline
  \hline
  0            & 1.3M      &  240k             & 330k          &   290k              & 290k          &   180k     \\
  1            & 1.4M      &  220k             & 350k          &   260k              & 320k          &   200k     \\
  2            & 1.3M      &  250k             & 330k          &   300k              & 270k          &   160k     \\
  3            & 1.3M      &  280k             & 360k          &   270k              & 250k          &   120k     \\
  4            & 1.2M      &  300k             & 370k          &   250k              & 230k          &   90k       \\
  5            & 1.2M      &  370k             & 330k          &   270k              & 190k          &   50k       \\
  \end{tabular}
  \end{center}
  \label{tab-eventrate} 
  \end{table}

In addition to the position scan for physics, some surveys for the detector response, gain, threshold, trigger
logic, will be performed.
It is also planned to repeat short operations at position 0 to monitor the `zero' degree position.
Including these surveys we require one day of RHICf physics operation. 
For contingency, one more day of operation time is requested.


\chapter{Expected results} \label{sec-result}
\section{Single particle spectra} \label{sec-spectrum}
With 2\,hour of data taking a dataset corresponding to an effective integrated luminosity of
2\,nb$^{-1}$ or 1$\times$10$^{8}$ inelastic collisions is obtained for single shower events.
Expected photon and neutron spectra with 10$^{8}$ collisions taken at each of 6 scan positions (12\,hours) are shown in 
Fig.\ref{fig-photon-spectra} and Fig.\ref{fig-neutron-spectra}, respectively.
The results for 4 rapidity bins 6.26$<\eta<$6.49, 6.87$<\eta<$7.40,  7.40$<\eta<$7.83 and 8.27$<\eta$ are presented
as examples.
In the calculation, event generators PYTHIA 8.185 \cite{PYTHIA}, EPOS LHC \cite{EPOS} and QGSJET II-03 \cite{QGS}
are used.
Only statistical errors are indicated in the plots.
Neutron spectra after taking into account the 70\% detection efficiency, 40\% energy resolution and 1\,mm position 
resolution are shown in Fig.\ref{fig-neutron-spectra2}.
A possible improvement of the energy resolution by a joint analysis with the ZDC is not considered in these plots.
In any cases, except at the highest energy bins, statistical errors are small enough that the differences between models 
will be clearly evident. 
Since the uncertainty will be dominated by possible systematic errors, a careful studies of detector performance and beam
condition are necessary.
An advantage of the RHICf measurements is that the performance of the detector have been studied at the CERN SPS
beam line using 100-350\,GeV proton \cite{LHCf-neutron-performance} and 100-250\,GeV electron 
\cite{LHCf-EM-performance} beams coinciding with the energy range at RHIC.

  \begin{figure}[h]
  \begin{center}
  \includegraphics[width=6cm]{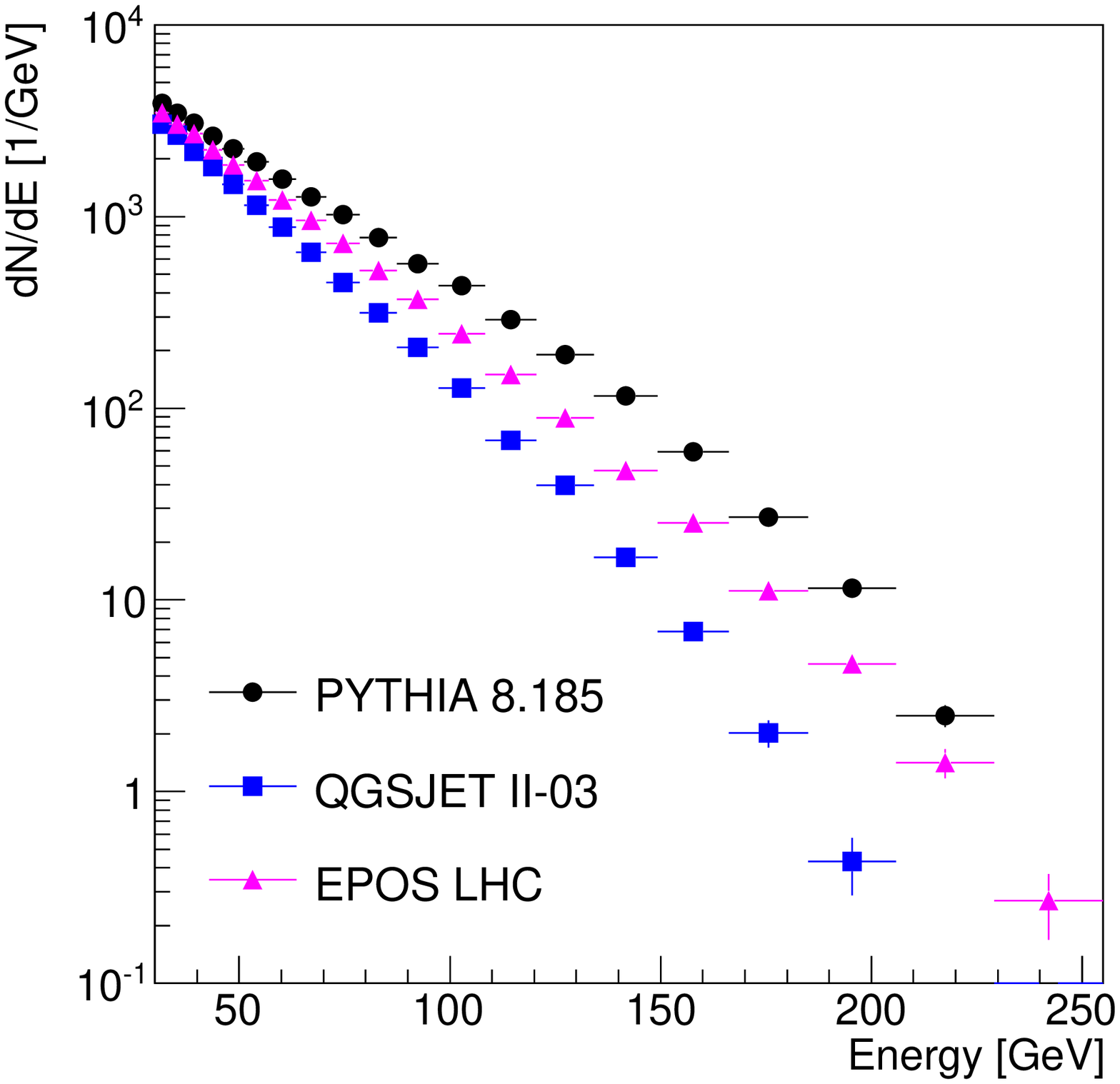}
  \includegraphics[width=6cm]{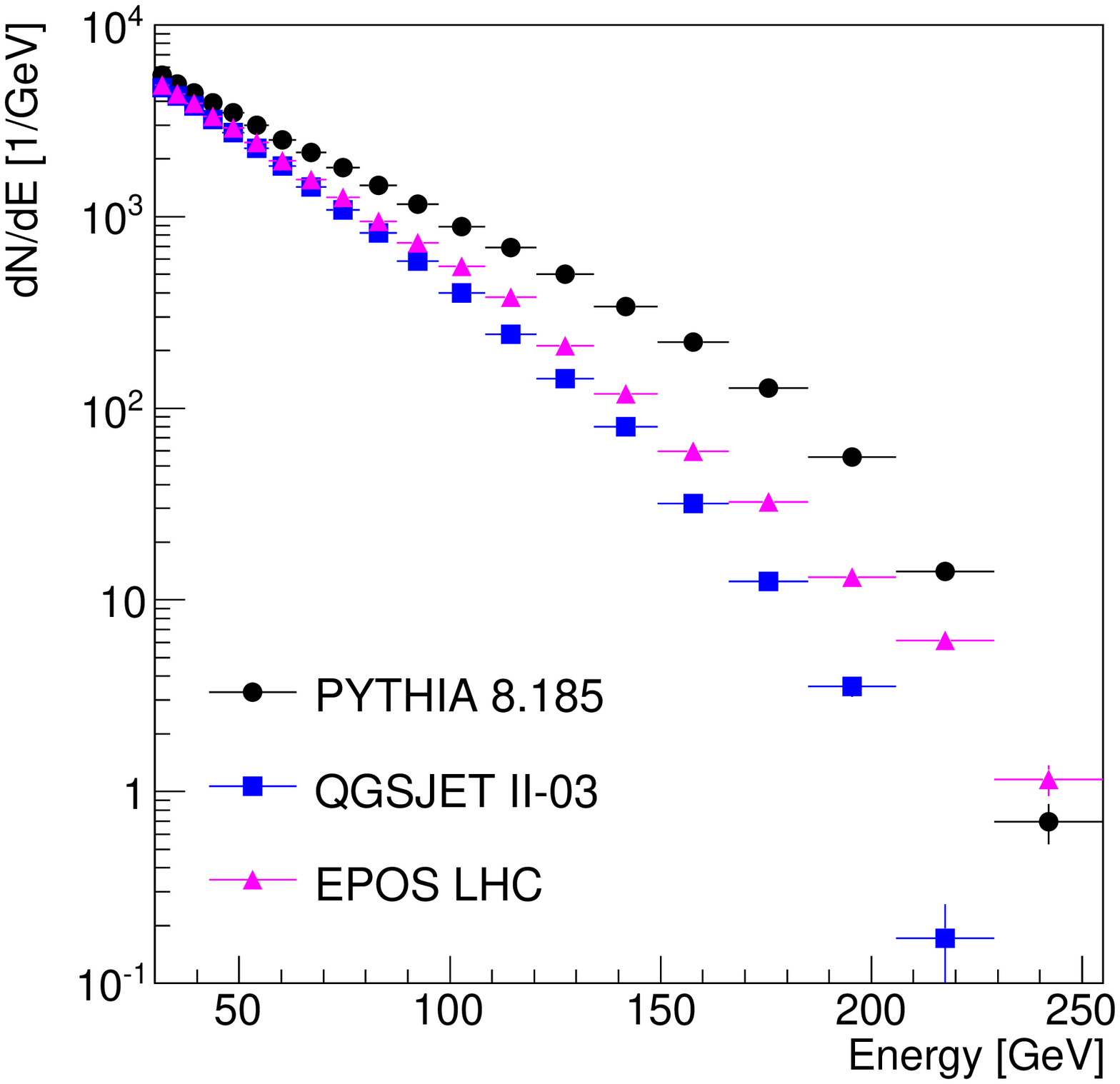}
  \includegraphics[width=6cm]{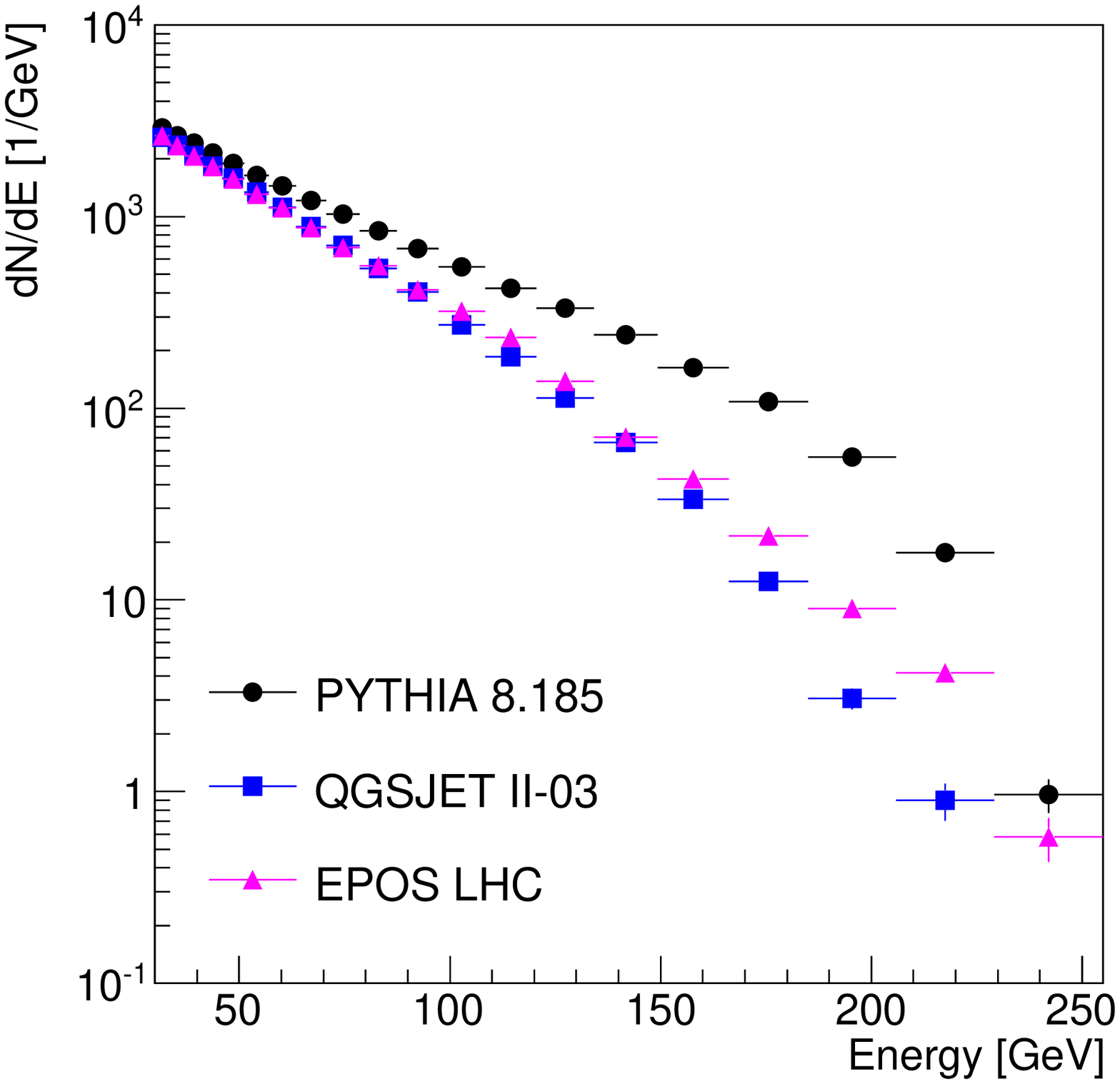}
  \includegraphics[width=6cm]{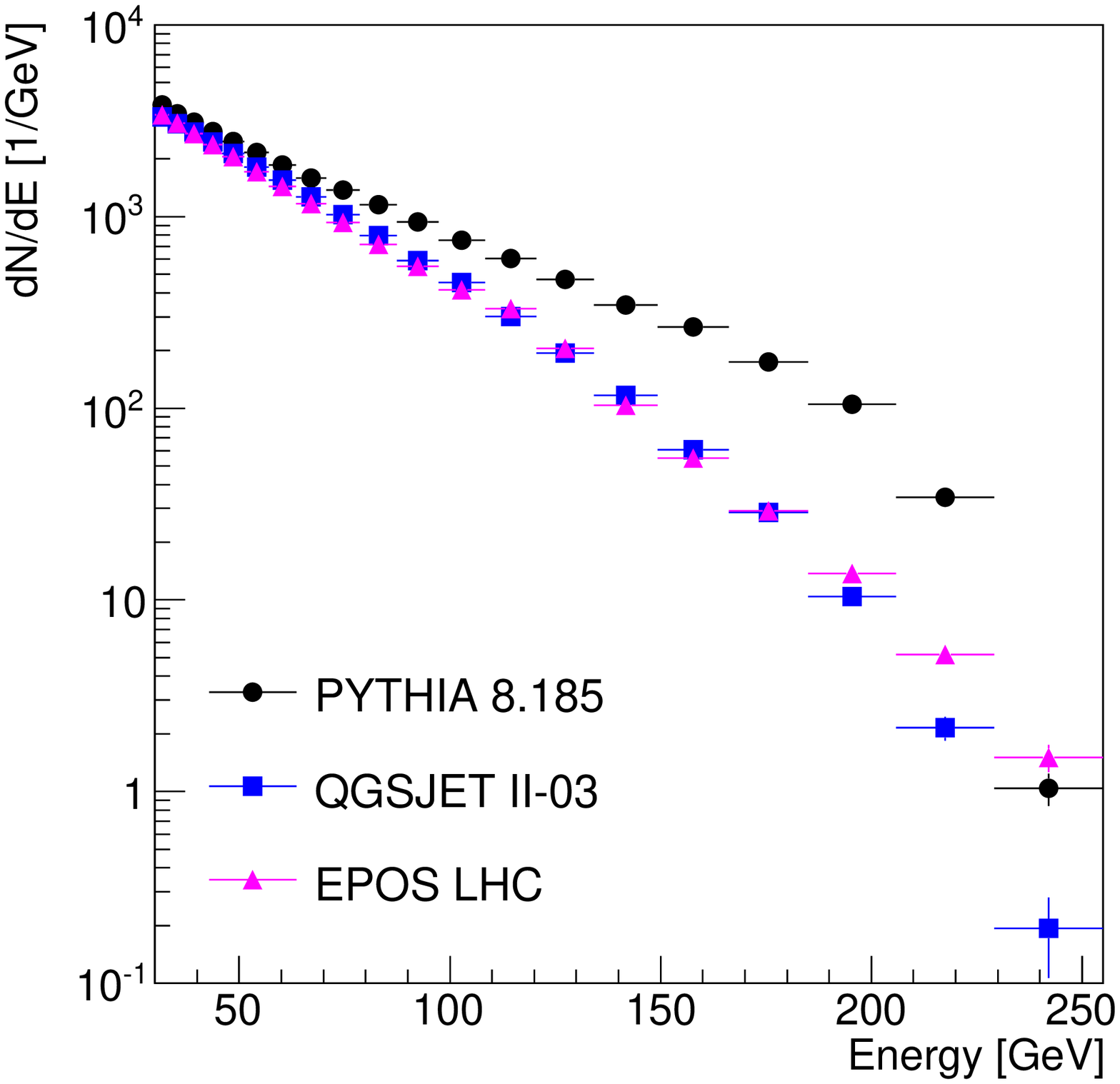}
  \caption{Energy spectra of photons expected from a 2\,hours$\times$6 positions dataset at  6.26$<\eta<$6.49 (top-left), 6.87$<\eta<$7.40 (top-right) 7.40$<\eta<$7.83 (bottom-left) and 8.27$<\eta$ (bottom-right).    Different colors designate event generators used in the calculation.}
  \label{fig-photon-spectra}
  \end{center} 
  \end{figure}

  \begin{figure}[h]
  \begin{center}
  \includegraphics[width=6cm]{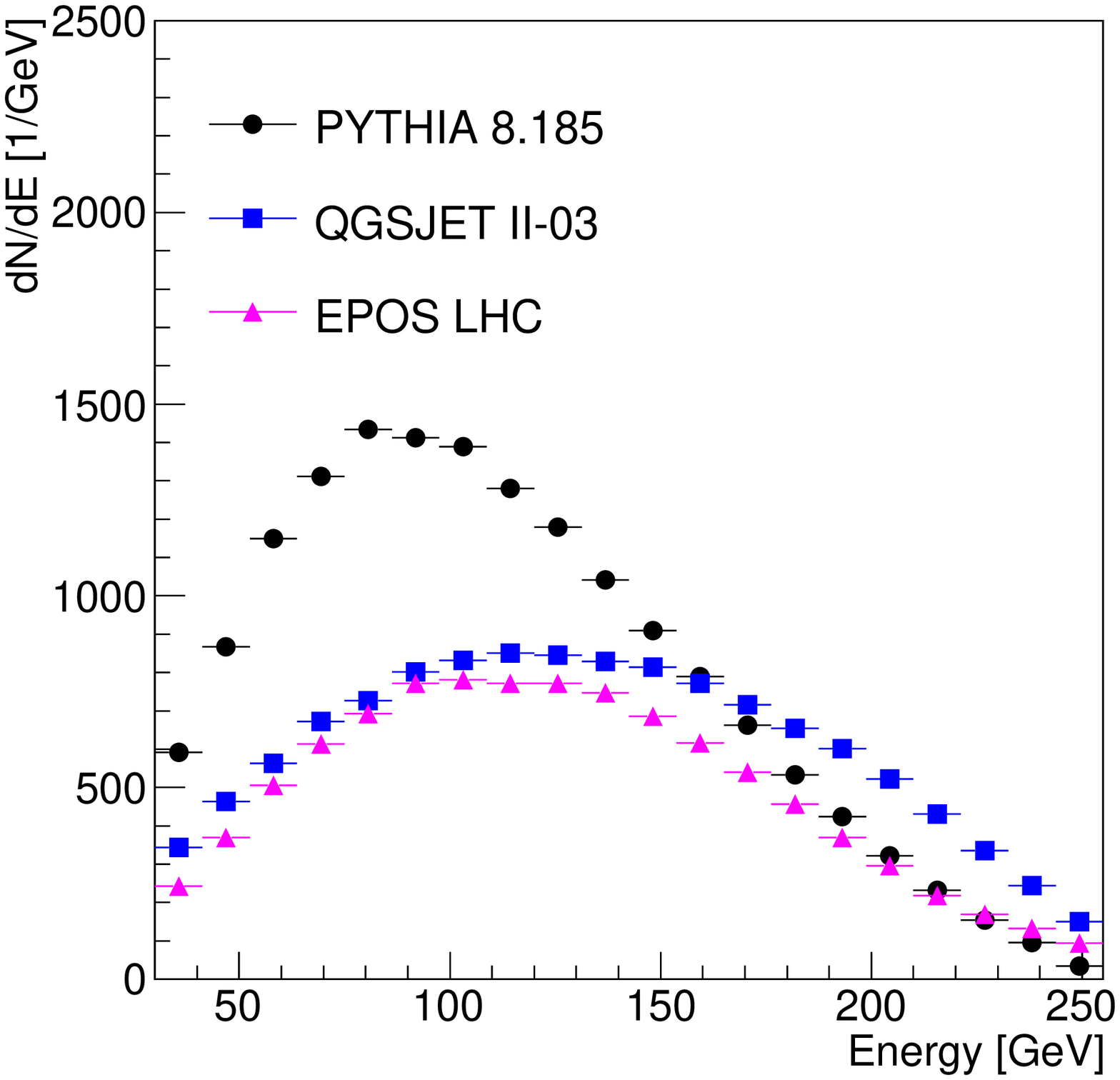}
  \includegraphics[width=6cm]{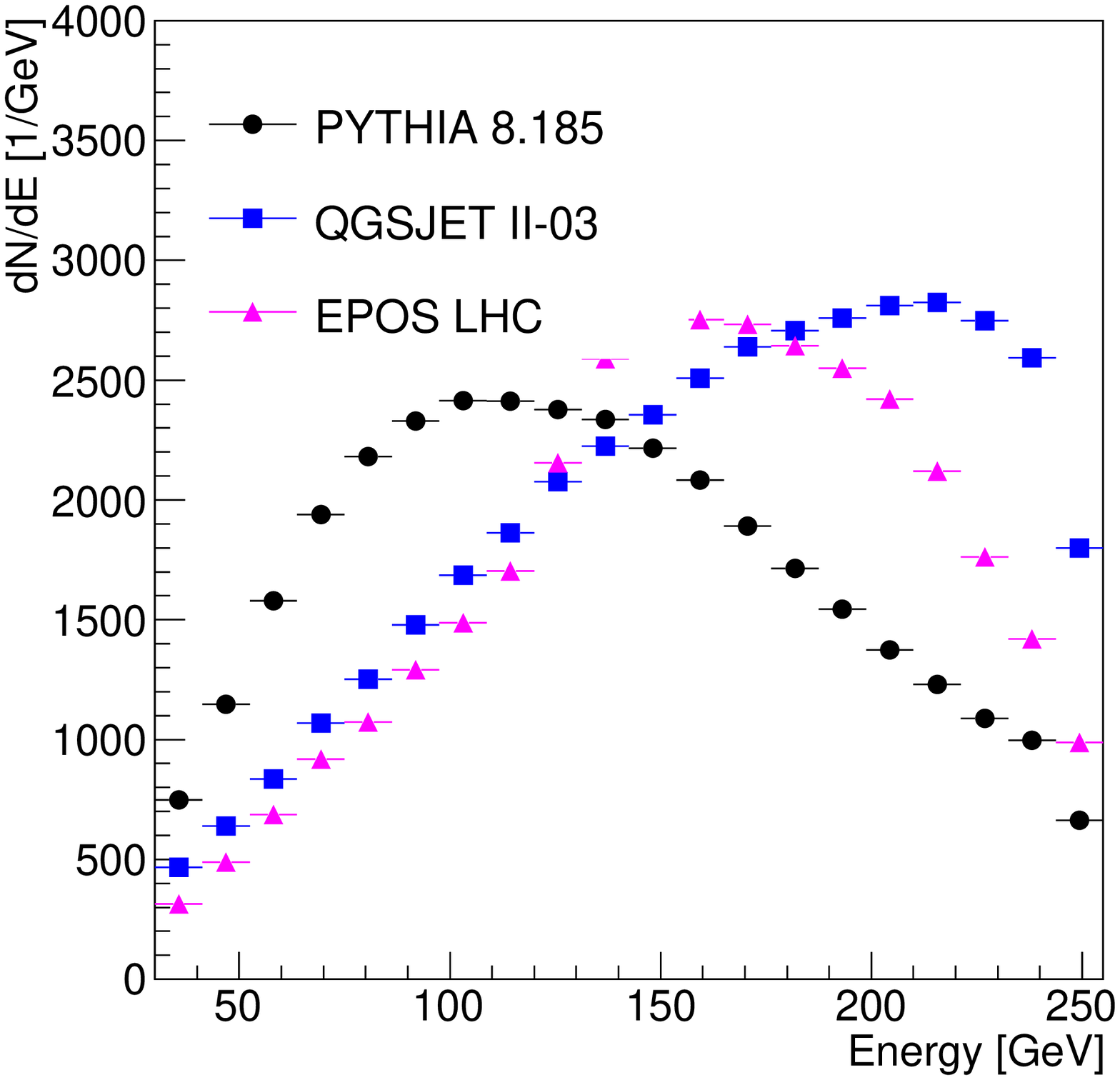}
  \includegraphics[width=6cm]{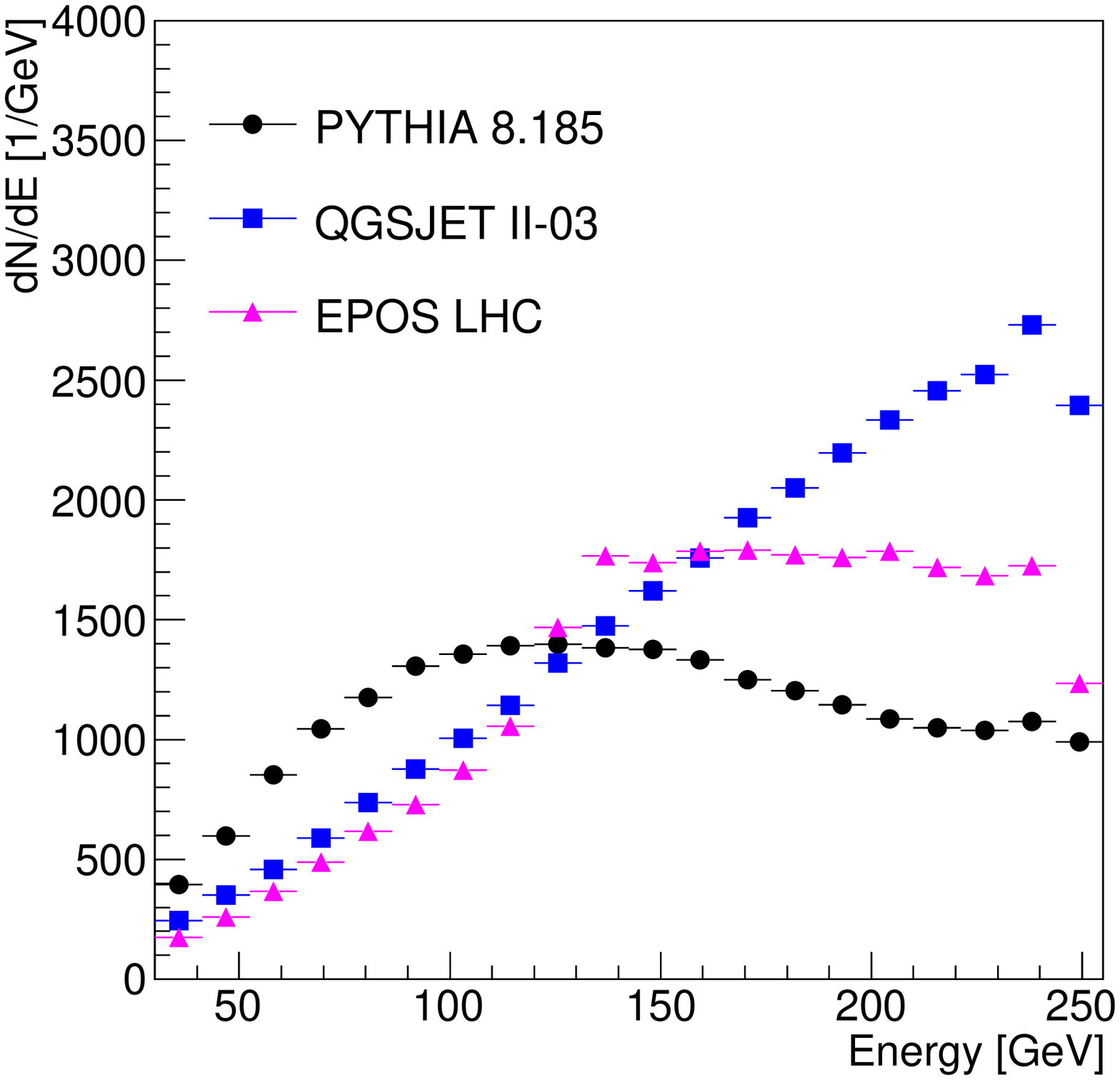}
  \includegraphics[width=6cm]{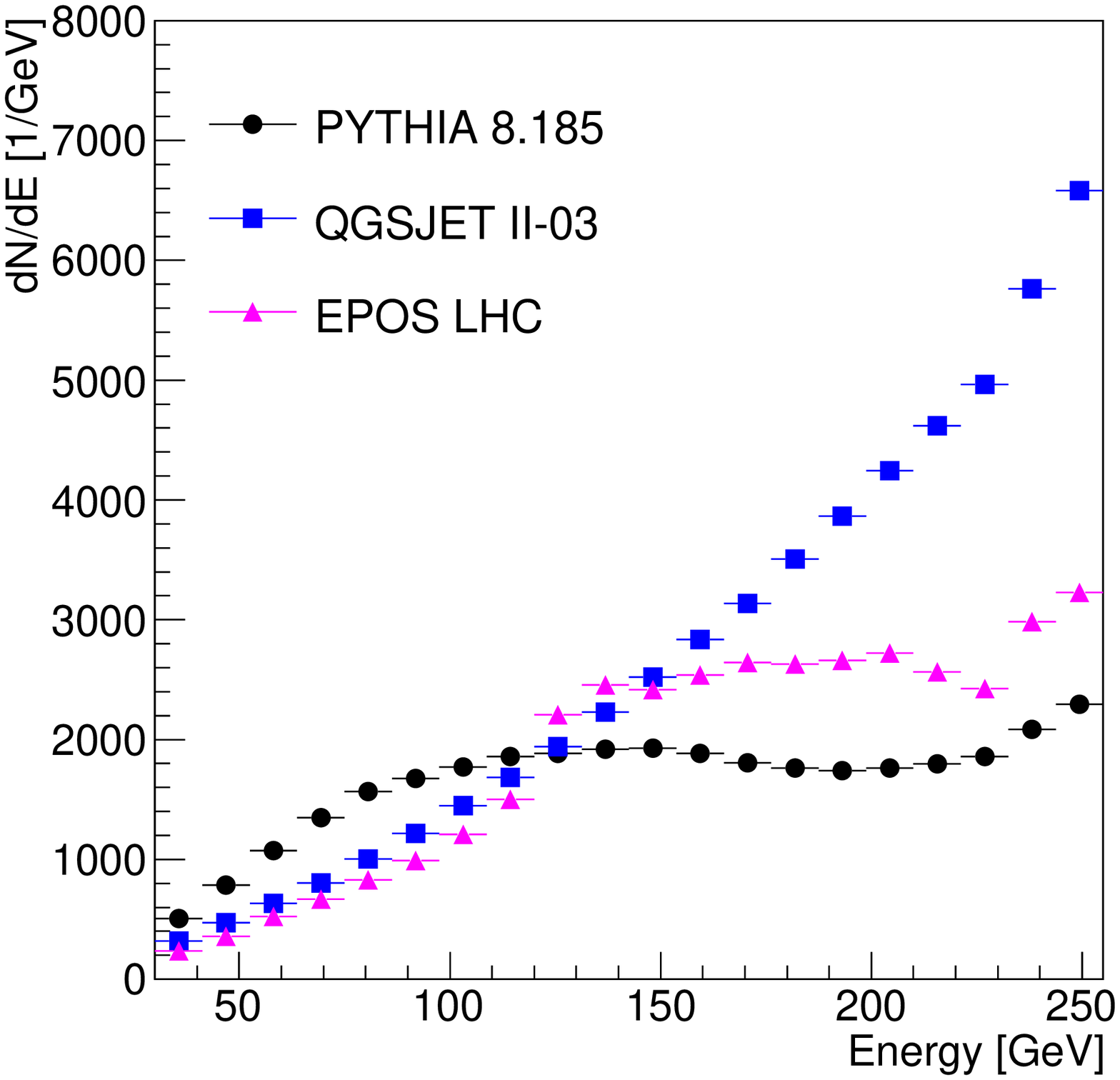}
  \caption{Energy spectra of neutrons expected from a 2\,hours$\times$6 positions dataset at  6.26$<\eta<$6.49 (top-left), 6.87$<\eta<$7.40 (top-right) 7.40$<\eta<$7.83 (bottom-left) and 8.27$<\eta$ (bottom-right).    Different colors designate event generators used in the calculation.}
  \label{fig-neutron-spectra}
  \end{center} 
  \end{figure}

  \begin{figure}[h]
  \begin{center}
  \includegraphics[width=6cm]{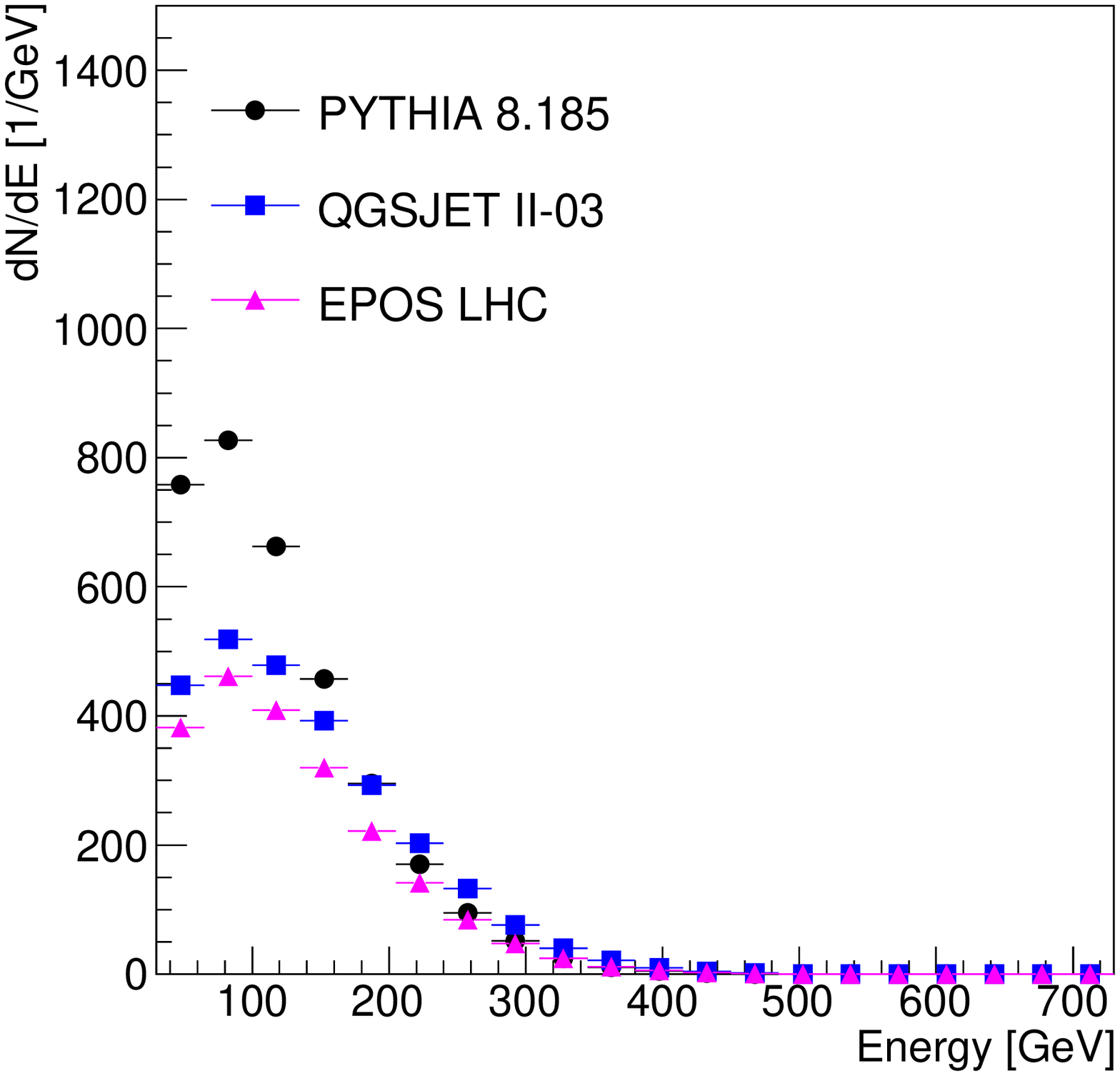}
  \includegraphics[width=6cm]{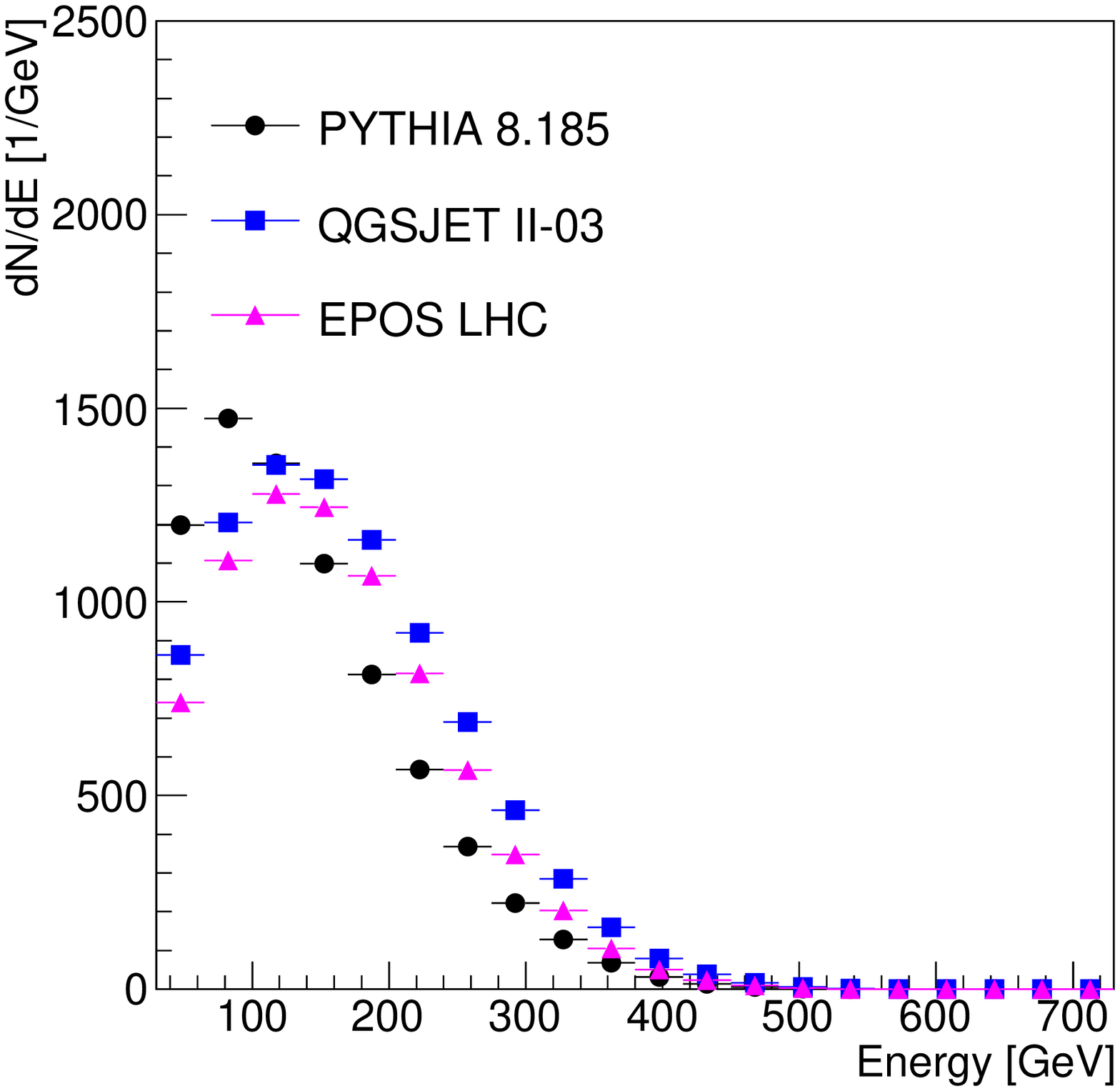}
  \includegraphics[width=6cm]{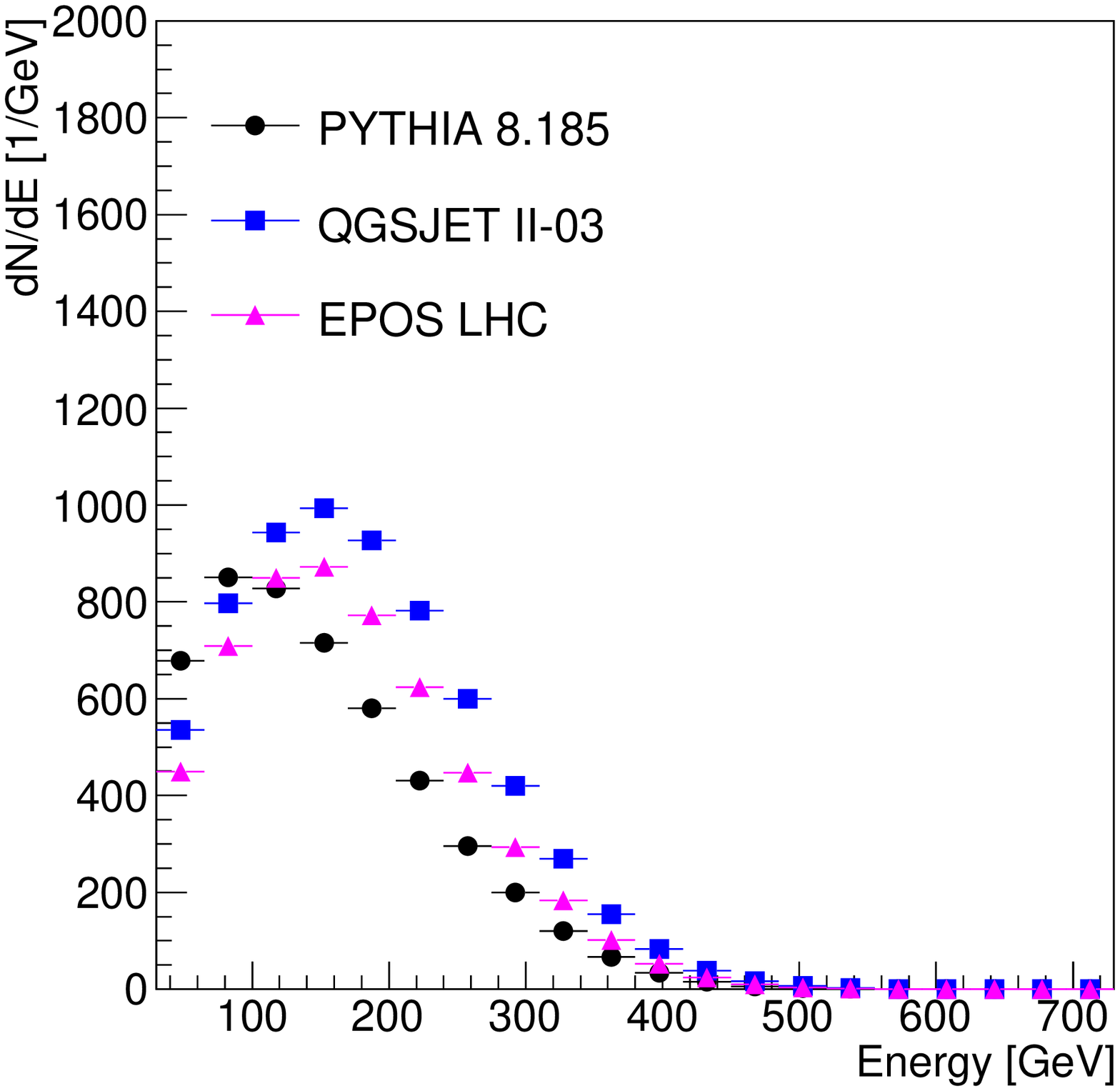}
  \includegraphics[width=6cm]{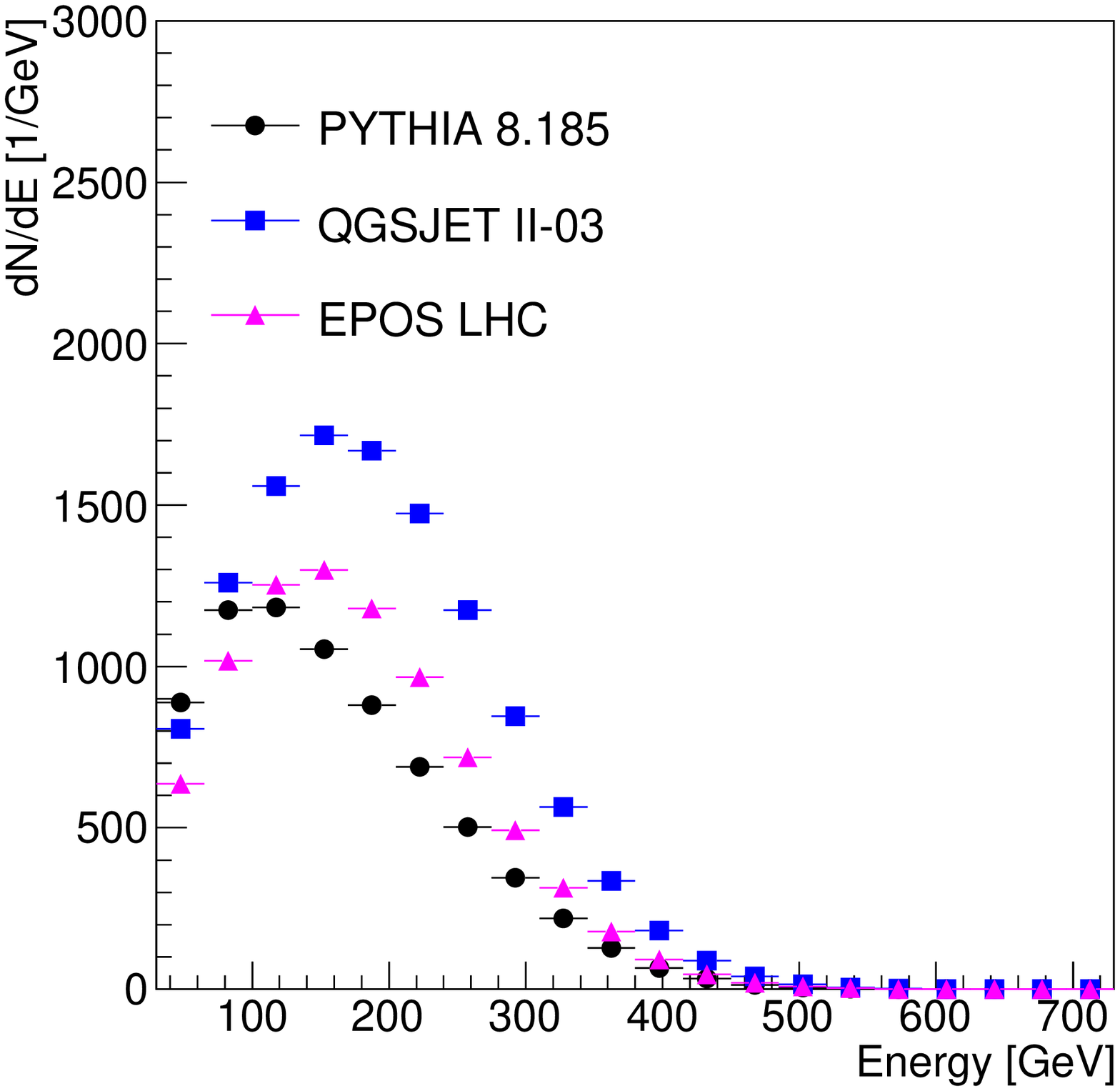}
  \caption{Energy spectra of neutrons expected from a 2\,hours$\times$6 positions dataset at  6.26$<\eta<$6.49 (top-left), 6.87$<\eta<$7.40 (top-right) 7.40$<\eta<$7.83 (bottom-left) and 8.27$<\eta$ (bottom-right).   Effects of detection efficiency, energy resolution and position resolution of the RHICf detector are taken into account.  Different colors designate event generators used in the calculation.}
  \label{fig-neutron-spectra2}
  \end{center} 
  \end{figure}

\section{$\pi^{0}$ spectra}
With 30\,min of data taking, a dataset corresponding to an effective integrated luminosity 
16\,nb$^{-1}$ or 8$\times$10$^{8}$ inelastic collisions is obtained.
Expected $\pi^{0}$ spectra with 10$^{8}$ collisions, corresponding to 4\,min (this is limited by the CPU time in simulation),
at 6.04$<$y$<$6.16, 6.36$<$y$<$6.70, 6.70$<$y$<$6.88 and 7.25$<$y$<$7.62 are shown in 
Fig.\ref{fig-pi0-spectra} where y is rapidity of $\pi^{0}$.
In the calculation, event generators PYTHIA 8.185, EPOS LHC and QGSJET II-03 are used.
Only statistical errors are indicated in the plots.
With this MC statistics, the statistical errors are at the 10\% level for each bin.
With a reasonable data taking time, we can expect the statistical errors to be well below the systematic uncertainties.
The differences between models are clearly evident already with the statistics obtained from the short data taking. 

  \begin{figure}[h]
  \begin{center}
  \includegraphics[width=6cm]{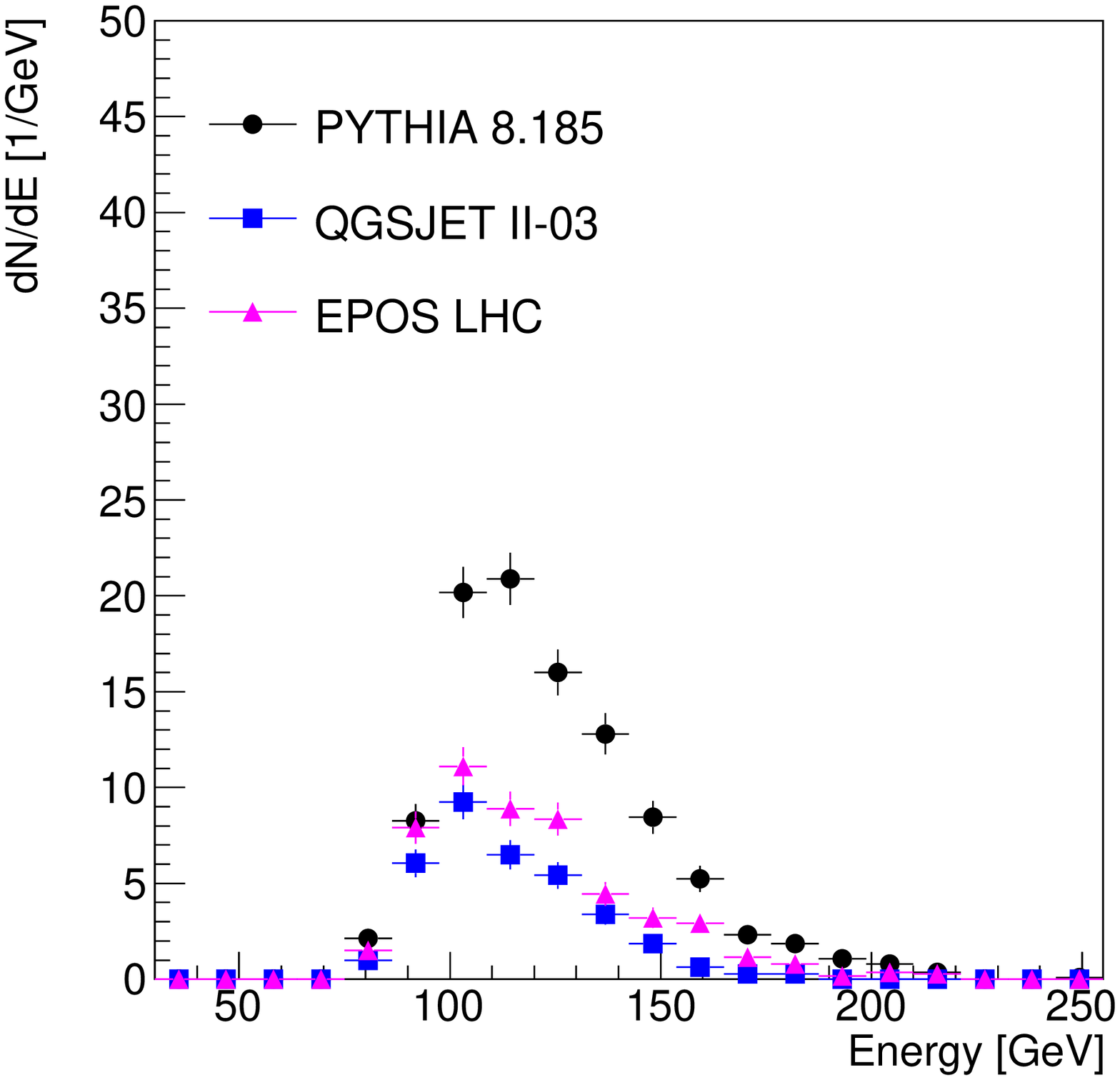}
  \includegraphics[width=6cm]{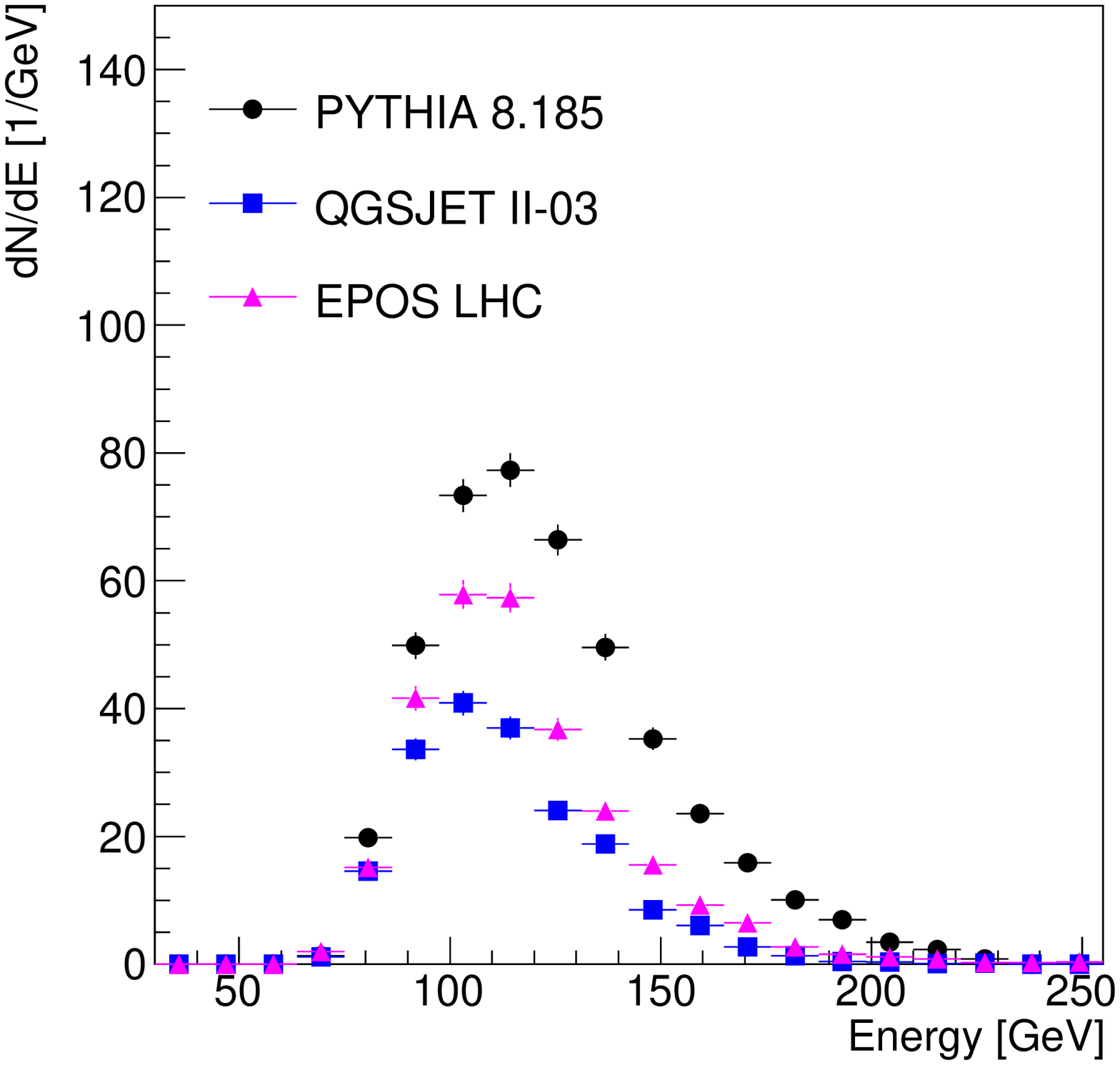}
  \includegraphics[width=6cm]{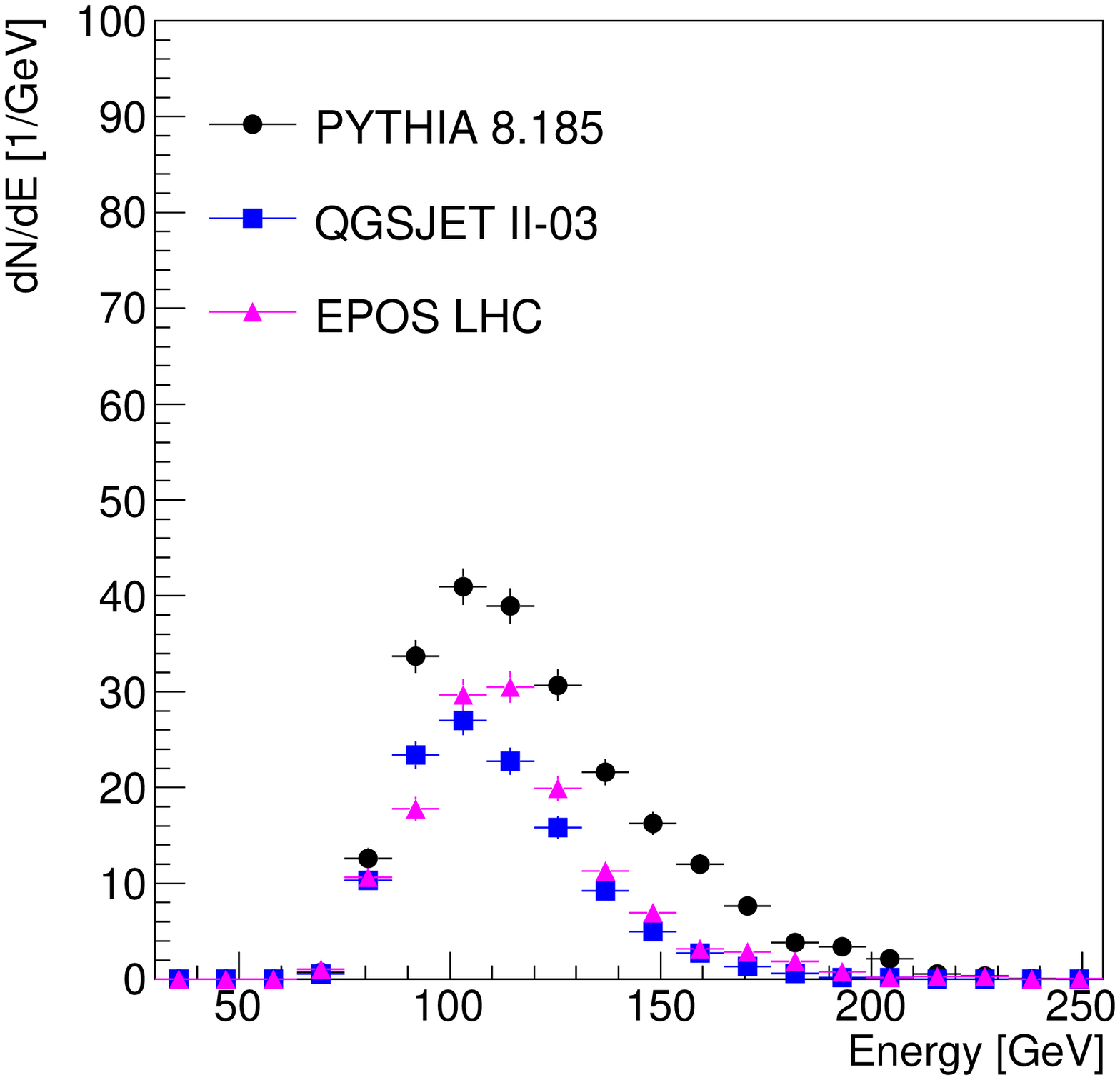}
  \includegraphics[width=6cm]{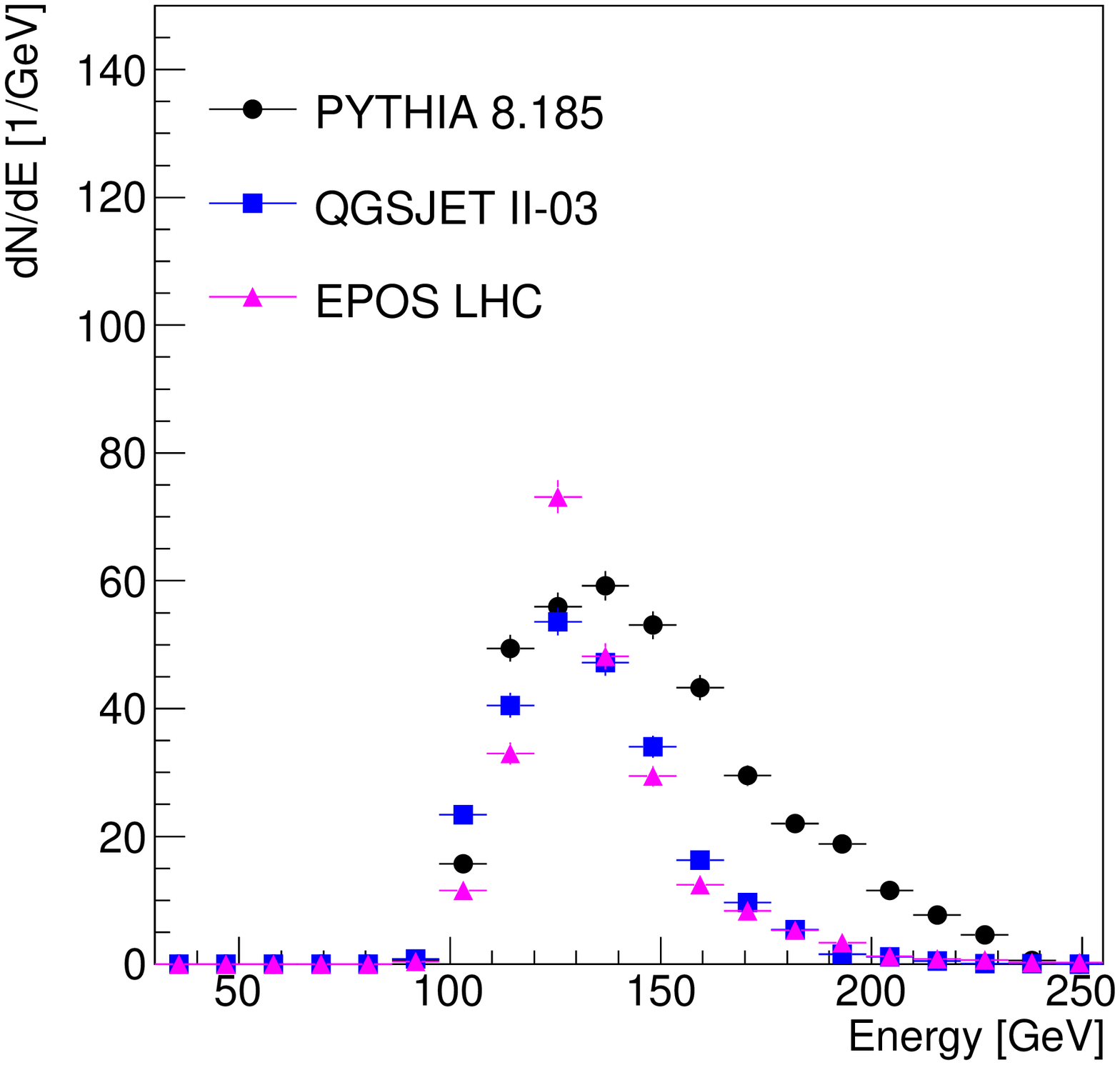}
  \caption{Energy spectra of $\pi^{0}$ expected from a 4\,min$\times$6 positions dataset at  6.04$<$y$<$6.16 (top-left), 6.36$<$y$<$6.70 (top-right) 6.70$<$y$<$6.88 (bottom-left) and 7.25$<$y$<$7.62 (bottom-right).     Different colors designate event generators used in the calculation.}
  \label{fig-pi0-spectra}
  \end{center} 
  \end{figure}

\section{Spin asymmetry}
Using the same data set to the spectrum analysis, RHICf can study the spin asymmetry like PHENIX but with a better 
position resolution and hence a better p$_{T}$ resolution than the PHENIX SMD.
The vertical scan allows RHICf to cover up to higher p$_{T}$ than PHENIX.
Expected numbers of events with x$_{F}>$0.4 in several p$_{T}$ bins are  summarized in Tab.\ref{tab-asymstat}. 
Effective number of collisions (luminosity) of 10$^{8}$ (2\,nb$^{-1}$) and 10$^{9}$ (20\,nb$^{-1}$) at each of 6 positions are
assumed for the single shower events (neutrons and photons) and $\pi^{0}$ events, respectively. 
These correspond to a data taking time of 12 and 4\,hours, respectively, and can be completed during the spectral
measurements discussed in Sec.\ref{sec-spectrum}.
Statistical accuracies for determining the amplitude of asymmetry ($\delta$A) are also summarized in the table.
Assuming a polarization P to be 50\%, $\delta$A is defined as $1/(P \sqrt{N})$.
According to these statistics, $\sim$1\% statistical accuracy is obtained at p$_{T}<$1.0\,GeV/c, 0.5\,GeV/c and 0.5\,GeV/c 
for neutrons, photons and $\pi^{0}$, respectively.
These extend the past PHENIX measurements with good overlapping p$_{T}$ coverages.
Expected data points given by RHICf overlaid on the past PHENIX result are shown in Fig.\ref{fig:asym_pt_mod} as
red ellipses.
Here the sizes of the ellipses indicate the expected p$_{T}$ resolution of RHICf \cite{RHICf-LOI} and 
$\pm$1\% errors on asymmetry.

There are some options under consideration for the asymmetry measurements.
\begin{itemize}
  \item High energy enhanced trigger to increase the statistics of high energy (high p$_{T}$) events.
  \item Trigger using the PHENIX Beam Beam Counter (BBC) as was done in the PHENIX analysis.
\end{itemize}


  \begin{table}[htbp]
  \begin{center}
  \caption{Statistics (1,000 events) obtained from 12\,nb$^{-1}$ (12\,hours) and 120\,nb$^{-1}$ (4\,hours) effective luminosities for single shower events (neutron and photon) and $\pi^{0}$ events, respectively with x$_{F}>$0.4.  
  $\delta$A indicates the expected statistical accuracy of the asymmetry determination.}
  \vskip 3mm
  \begin{tabular}{lcccccc}
                  & \multicolumn{2}{c}{neutron} & \multicolumn{2}{c}{photon}  & \multicolumn{2}{c}{$\pi^0$} \\
  $p_T$ (GeV/$c$) & $N (\times10^{3})$  & $\delta A$ & $N (\times10^{3})$ & $\delta A$ & $N (\times10^{3})$ & $\delta A$ \\
  \hline
  \hline
  0.0 -- 0.1      &     660  & 0.0025           &     110     & 0.0060           &   100   & 0.0063 \\
  0.1 -- 0.2      &     920  & 0.0021           &     120     & 0.0058           &   130   & 0.0055 \\
  0.2 -- 0.3      &     820  & 0.0022           &     110     & 0.0060           &     89   & 0.0067 \\
  0.3 -- 0.4      &     670  & 0.0024           &       79     & 0.0071           &     58   & 0.0083 \\
  0.4 -- 0.5      &     450  & 0.0030           &       43     & 0.0096           &     37   & 0.010 \\
  0.5 -- 0.6      &     250  & 0.0040           &       18     & 0.015             &     14   & 0.017 \\
  0.6 -- 0.8      &     170  & 0.0049           &         8     & 0.022             &       8   & 0.022 \\
  0.8 -- 1.0      &       29  & 0.012             &         1     & 0.063             &       1   & 0.063 \\
  \end{tabular}
  \end{center}
  \label{tab-asymstat} 
  \end{table}

  \begin{figure}[htbp]
  \begin{center}
  \includegraphics[width=10cm]{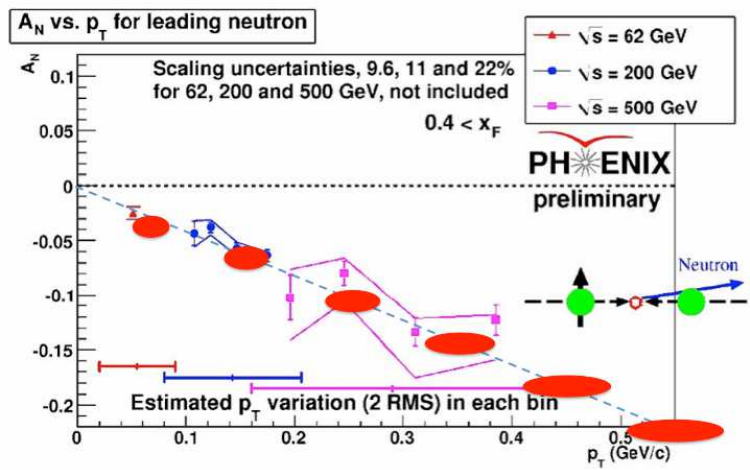}
  \caption{Expected RHICf result for neutron asymmetry is plotted as red ellipses on the past PHENIX result. RHICf p$_{T}$ resolution and $\pm$1\% errors are indicated by the size of the ellipses.}
  \label{fig:asym_pt_mod}
  \end{center} 
  \end{figure}


\chapter{Schedule and expected supports from BNL} \label{sec-schedule}
\section {Schedule}
\noindent {\bf Detector}\\
\\
LHCf will finish its operation at LHC in early May 2015.
Then the detector will be removed from the tunnel during a technical stop in June.
Because a weak activation is expected at the time of removal, a few weeks of cooling at the CERN site
is necessary.
The arrival of the detector at BNL will be around October. 
After a checkout of the detector condition at BNL, it will be installed in RHIC.

\vskip 8mm
\noindent {\bf DAQ and mechanical setup}\\
\\
Before the detector installation in 2015 autumn, we need to perform
\begin{itemize}
  \item Cabling from RHICf installation slot to the rack room
  \item Construction of the detector support structure 
  \item Mockup test for detector installation
  \item Setup of the electronics in the rack room
  \item Setup of the counting room
  \item Dry run of the data taking, synchronization with accelerator and PHENIX 
\end{itemize}

  \begin{figure}[h]
  \begin{center}
  \includegraphics[width=14cm]{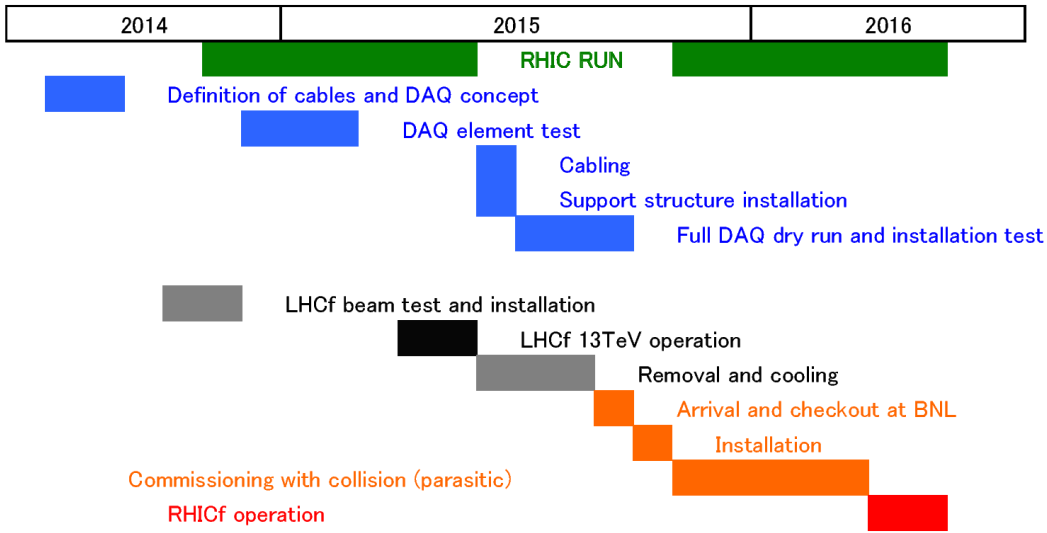}
  \caption{Schedule until RHICf operation}
  \label{fig-timeline}
  \end{center} 
  \end{figure}

A time line of the preparation is shown in Fig.\ref{fig-timeline}.
Grey and black items indicate the major activities related to LHCf.
Blue items show the activities carried out without the detector that can be proceeded immediately after
the approval of this proposal.
The schedule of the detector transport after the LHCf operation is drawn by orange and the RHICf
physics operation is indicated by red.
The timing of the final operation depends on the actual scheduling of RUN16, but here the operation
at the end of RUN16 is supposed.

\section{Expected supports from BNL}
Some technical supports including man power from BNL and/or PHENIX are expected to prepare the 
experiment.
They are listed up below.

\begin{itemize}
  \item Cabling from the RHICf installation slot to the PHENIX rack room
  \item Construction and installation of the detector support structure
  \item Transportation, installation and geometrical survey of the detector
  \item Support for custom process not to delay the detector arrival from CERN
\end{itemize}

The cost is supposed to be covered by the RHICf collaboration. 

\chapter{Manpower and budget} \label{sec-budget}
The project is carried out mainly by Nagoya university, Waseda university and RIKEN in Japan and 
University of Florence, University of Catania and INFN in Italy.
Addition to the authors in this proposal, some graduate students will participate to this experiment to write
their PhD theses.
All members including students are trained enough in the PHENIX and the LHCf experiments to carry out
the RHICf project efficiently.

Because the detector and the major electronics are already available, not large amount of budget is required.
The major items are
\begin{itemize}
  \item Travel cost between Japan/Italy and BNL (approx. 100\,k USD)
  \item Updating the electronics (approx. 20\,k USD)
  \item Cost for cables and cabling work (approx. 10\,k USD)
  \item Construction of detector support (approx. 5\,k USD)
  \item Transport of the material from CERN to BNL (approx. 5\,k USD)
\end{itemize}
The cost can be manageable within the running budget of the member institutes and also grants now being
applied.
The approval of this proposal will help to obtain dedicated grants for the RHICf program.


\end{large}
\end{document}